\documentclass[a4paper,twocolumn,preprint,5p,authoryear,sort&compress,8pt]{elsarticle}

\usepackage{stfloats}
\usepackage{lipsum}
\journal{NeuroImage}
\usepackage{lineno,hyperref}
\usepackage{amssymb}
\usepackage{multicol}
\usepackage{wrapfig}

\usepackage{graphicx} 
\usepackage{filecontents}
\usepackage{bbold}
\usepackage{hyperref}
\usepackage{makeidx}  
\usepackage[misc]{ifsym}
\usepackage{amsmath,graphicx}
\usepackage{multirow}
\usepackage{array}
\usepackage{siunitx}
\usepackage{xcolor}

\DeclareMathOperator*{\argmin}{arg\,min}
\DeclareMathOperator{\Tr}{Tr}
\newcommand{\rpm}{\raisebox{.2ex}{$\scriptstyle\pm$}}
\modulolinenumbers[5]
\usepackage{mathtools}

\biboptions{longnamesfirst,semicolon}
\begin{document}

\begin{frontmatter}

\title{A Joint Network Optimization \\ Framework to Predict Clinical Severity from \\ Resting State Functional MRI Data}
\author[wse]{N.S.~D'Souza\corref{cor1}}
\ead{Shimona.Niharika.Dsouza@jhu.edu}
\author[KKI]{M.B.~Nebel}
\author[KKI]{N.~Wymbs}
\author[KKI,Neu,Ped]{S.H.~Mostofsky}
\author[wse]{A.~Venkataraman}

\cortext[cor1]{Corresponding author}

\address[wse]{Department of Electrical and Computer Engineering, Johns Hopkins University, USA}
\address[KKI]{Center for Neurodevelopmental \& Imaging Research, Kennedy Krieger Institute}
\address[Neu]{Department of Neurology, Johns Hopkins School of Medicine, USA}
\address[Ped]{Department of Pediatrics, Johns Hopkins School of Medicine, USA}

\begin{abstract}
We propose a novel optimization framework to predict clinical severity from resting state fMRI (rs-fMRI) data. Our model consists of two coupled terms. The first term decomposes the correlation matrices into a sparse set of representative subnetworks that define a network manifold. These subnetworks are modeled as rank-one outer-products which correspond to the elemental patterns of co-activation across the brain; the subnetworks are combined via patient-specific non-negative coefficients. The second term is a linear regression model that uses the patient-specific coefficients to predict a measure of clinical severity. We validate our framework on two separate datasets in a ten fold cross validation setting. The first is a cohort of fifty-eight patients diagnosed with Autism Spectrum Disorder (ASD). The second dataset consists of sixty three patients from a publicly available ASD database. Our method outperforms standard semi-supervised frameworks, which employ conventional graph theoretic and statistical representation learning techniques to relate the rs-fMRI correlations to behavior. In contrast, our joint network optimization framework exploits the structure of the rs-fMRI correlation matrices to simultaneously capture group level effects and patient heterogeneity. Finally, we demonstrate that our proposed framework robustly identifies clinically relevant networks characteristic of ASD. 
\end{abstract}


\end{frontmatter}

\section{Introduction}
\begin{figure*}[t!]
   \centering
   \includegraphics[scale=0.5]{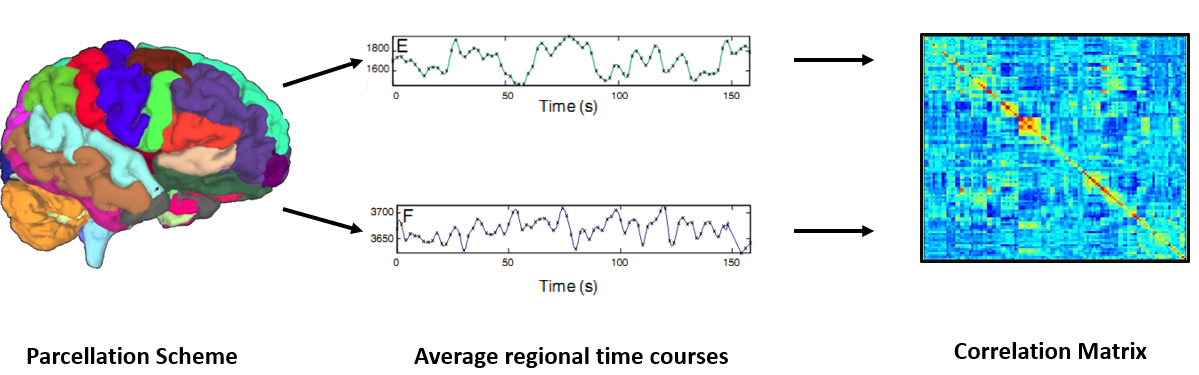}
   \caption{We group voxels in the brain into ROIs defined by a standard atlas and compute the average time courses for each ROI. The correlation matrix captures the synchrony in the average time courses}\label{Ex}
\end{figure*}
{R}{esting} State fMRI (rs-fMRI) is a non-invasive neuroimaging modality that captures steady-state patterns of co-activation in the brain in the absence of a task paradigm. It is believed that these correlation patterns reflect the intrinsic communication between brain regions [\cite{fox2007spontaneous}]. Consequently, rs-fMRI has become ubiquitous in the characterization of neuropsychiatric disorders such as Autism Spectrum Disorder (ASD) [\cite{minshew2010nature}], Attention Deficit Hyperactivity Disorder (ADHD) [\cite{bush2005functional}], and schizophrenia [\cite{niznikiewicz2003recent}]. Traditional rs-fMRI analysis has concentrated on comparing the statistics of the rs-fMRI data, or variations in these statistics, across individuals or between different cohorts. For example, statistical differences in rs-fMRI features between a patient cohort and neurotypical controls have been considered as biomarkers of a particular disorder. However, the high dimensionality of rs-fMRI data, along with the considerable inter-patient variability, make it extremely difficult to reliably predict clinical manifestations on a patient-specific basis. 
\par There has been considerable work in developing statistical methods to analyze rs-fMRI data. A large number of these studies build on standard multivariate [\cite{woo2017building}] or random effects models [\cite{holmes1998generalisability}] to capture population level differences in functional connectivity. Although these studies identify functional connections affected by the disease, they often fail to generalize on a patient-specific level. Additionally, these techniques do not adequately characterize distributed impairments across multiple brain systems, which is crucial for understanding the complex pathologies associated with neuropsychiatric disorders [\cite{kaiser2010neural,koshino2005functional,rippon2007disordered}]. This limitation has warranted the development of network-based models to study the inter and intra-subject variation across populations.
\par Network-based rs-fMRI studies typically group voxels in the brain into regions of interest (ROIs) using a standard anatomical or functional atlas. Further, the synchrony between the average regional time courses is summarized using a similarity matrix, which is the input for further analyses. This extraction procedure is demonstrated in Fig.~\ref{Ex}. The works of [\cite{sporns2004organization,rubinov2010complex,bullmore2009complex}] use graph theoretic notions of connectivity based on aggregate network measures, such as node degree, betweenness centrality, and eigenvector centrality to study the functional organization of the brain. These measures are extremely useful to compactly summarize the connectivity information onto a restricted set of nodes which map to brain regions. A more global network property is small-worldedness [\cite{bassett2006small}], which describes an architecture of sparsely connected clusters of nodes. Changes in small-worldedness have been implicated in many neurological disorders [\cite{liu2008disrupted},\cite{sanz2010loss}]. These characterizations are quite successful at capturing global connectivity information, but often fail to illuminate the underlying etiological mechanisms. 
\par To address the limitations of aggregate graph theoretic notions, recent focus has shifted towards mechanistic network models, which incorporate hierarchy onto existing graph connectivity notions. Community detection techniques are a class of population-level models which are used to identify highly interconnected subgraphs within a larger network. These techniques have become popular for understanding the organization of complex systems like the brain network architecture [\cite{bardella2016hierarchical}]. An application of this approach to identify regions having abnormal connectivity in schizophrenia patients can be found in [\cite{venkataraman2013connectivity}]. Similarly, Bayesian community detection algorithms developed in [\cite{venkataraman2016bayesian}] have provided valuable insights in characterizing the social and communicative deficits associated with autism. An alternative network topology is the hub-spoke model, which targets regions associated with a large number of altered rs-fMRI connections [\cite{venkataraman2013connectivity}, \cite{venkataraman2012brain}, \cite{venkataraman2015unbiased}]. However, the above methods focus on group characterizations, and even studies that consider patient variability [\cite{venkataraman2017unified}] have little generalization power on new subjects.  
\par Machine learning techniques cast the neuroimaging prediction problem as a two stage procedure. Essentially, the first step is a feature selection or a representation learning stage, while the second stage uses the output of the first to predict the subject characteristics. A simple representation learning framework entails a careful sub-selection of specialized biomarkers [\cite{ravishankar2016recursive,hong2017multidimensional}]. On a whole brain level, data-driven approaches treat the patient connectivity information as a \textit{feature map} and estimate lower dimensional projections, typically through PCA, kernel-PCA [\cite{sidhu2012kernel}] or ICA [\cite{uddin2013salience}]. From here, the most popular classifier (i.e. a stage two algorithm) is a Support Vector Machine (SVM) [\cite{ecker2010investigating}], which optimizes the decision boundary between patients and neurotypical controls [\cite{uddin2013salience}]. SVMs have also been shown to identify disease sub-types [\cite{hong2017multidimensional}] from the lower dimensional features with high accuracy. Along similar lines, the work of [\cite{hoyos2017frem}] proposes an ensemble learning based encoder-decoder model, that is able to provide competitive performance on large fMRI datasets at discriminative tasks. 
\par While this two stage pipeline has been successful in the classification realm, characterizing finer-grained measures of clinical severity in the fMRI literature has been restricted to associative analysis, as opposed to an actual prediction on unseen data. For example, the work of [\cite{nebel2016intrinsic}] identifies key visual and motor ICA components, which are then used to compute a visuo-motor measure that is significantly correlated with social-communicative and motor deficit measures in ASD. In the context of a continuous value prediction, [\cite{rahim2017joint}] develops a modified random forest regression algorithm for stacked multi-output score estimation from multiple ROI-voxel correlation maps. They demonstrate that it outperforms single score prediction. This strategy, however, does not permit a straightforward interpretation of the co-activation patterns explaining an individual severity score. Rather, it identifies regions that explain the complete set of scores jointly. Finally, deep learning methods have become popular for several neuroimaging data analysis. These models have the ability to efficiently learn complex abstractions of the input data without requiring careful feature engineering. As a result, they have been quite successful in a number of case/control classification tasks [\cite{plis2014deep}]. However, a downside to these models is the requirement of large amounts of training data for adequate generalization, which is rarely the case with clinical neuroimaging. Consequently, there has been limited success in predicting behavior from rs-fMRI data using neural networks. In summary, the unification of rs-fMRI and behavioral severity prediction, remains an open challenge.
\par Dictionary learning [\cite{batmanghelich2012generative,eavani2013unsupervised}] methods move away from the pipelined representations, and have recently gained traction due to their ability to simultaneously model both group level and patient specific information. The work of [\cite{eavani2015identifying}] proposed a correlation matrix decomposition strategy, in which, multiple rank one outer products capture an underlying \textit{generative} basis. The sparse basis representation identifies meaningful co-activation patterns common to all the patients, while patient-specific coefficients combine the subnetworks and model the individual variability in the dataset. An extension of their work [\cite{eavani2014discriminative}] looks at classification of young adults versus children, again, by the addition of an SVM like hinge loss. Our work builds on this representation by using the \textit{discriminative} nature of these coefficients to predict their clinical severity via a linear regression penalty. This Joint Network Optimization (JNO) framework combines both a generative and discriminative term, as opposed to a pipelined hyperparameter search. The generalizability of the model is indicated by the regression performance on unseen data, instead of the correlation fit as used in [\cite{eavani2015identifying}]. This refinement demonstrates the potential of our JNO framework in identifying patient-predictive biomarkers of a given disorder. 
\par We validate our framework on an rs-fMRI study of Autism Spectrum Disorder (ASD). Patients with ASD are known to manifest a wide spectrum of impairments in terms of social reciprocity, communicative functioning, and repetitive/restrictive behaviours [\cite{spitzer1980diagnostic}]. This variation is typically quantified by a clinical severity measure obtained from an expert assessment. We find that our method outperforms several graph theoretic and machine learning feature representation techniques in predicting these severity scores on unseen data. Additionally, our model automatically extracts key networks commonly associated with altered functioning in the ASD literature. Finally, we quantify the merit of our joint objective function by comparing the combined predictive performance with that of a similarly defined two-stage decomposition and regression. We demonstrate that the joint objective bridges the gap in the two representative views of the data, thus aiding the resting state ASD characterization.
\begin{figure*}[t!]
   \centering
   \includegraphics[width= \textwidth- 2cm]{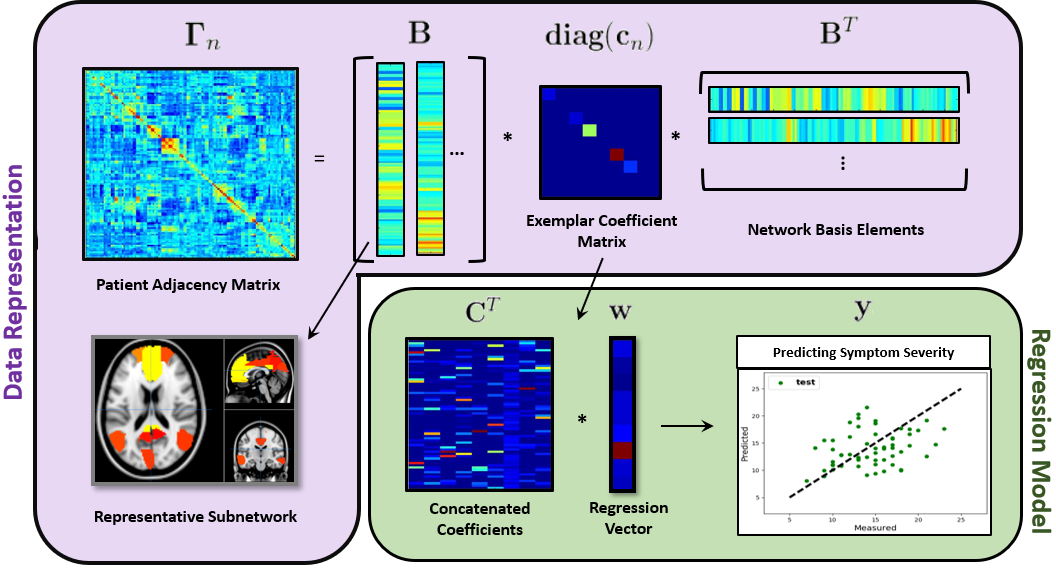}
   \caption{A two level joint model for connectivity and prediction. \textbf{Purple Box:} Depicts the functional data representation or `generative' term. The correlation matrix is decomposed into a group basis term and a patient specific coefficient term. The columns of the basis matrix correspond to individual subnetworks when projected onto the brain. We stack these coefficients into a matrix. \textbf{Green Box:} Prediction of symptom severity via linear regression } \label{fig:Fig1}
\end{figure*}
\par A preliminary version of our work appeared in [\cite{d2018generative}]. Here, we provide a detailed analysis of our model. We demonstrate the predictive performance of our algorithm on three different clinical severity measures, which capture varied social, behavioral and cognitive deficits associated with ASD. We evaluate our model on two clinical datasets to demonstrate the~reproducibility of the method. We identify resting state networks explaining the different behavioral manifestations, and accordingly discuss the robustness of our brain basis characterization across behavioral measures. Lastly, we study hyperparameter sensitivity, generalizability in a test-retest setting and include mitigation strategies to improve the robustness of the framework. 

\section{Materials and Methods }

\subsection{A Joint Model for Connectomics and Clinical Severity}
Fig.~\ref{fig:Fig1} presents a graphical overview of our model. The two inputs to our model are the rs-fMRI similarity matrices (upper left) and the scalar clinical severity scores for each patient (lower right). As mentioned earlier, Fig.~\ref{Ex} illustrates the construction of the similarity matrix from the data. These matrices quantify the Pearson's Correlation Coefficient between the average time courses for each region of interest (ROI). The clinical scores are obtained from an expert evaluation and quantify the severity of the symptoms for the individual.
\par Notice that the correlation matrices in Fig.~\ref{fig:Fig1} have a dual representation. The generative part of the model is indicated in the purple box. Here, we decompose the correlation matrix into a basis term and a patient coefficient term. The columns of the basis capture ROI co-activation patterns common to the entire cohort, while the coefficients differ across patients and quantify the strength of each basis column in the matrix representation. The green box indicates the discriminative part of the model. Here, we leverage the information from the patient-specific coefficients to estimate a given measure of clinical severity via a linear regression model for each individual.

\paragraph{\textbf{Rs-fMRI Data Representation}} We define $\mathbf{\Gamma}_{n} \in \mathcal{R}^{P \times P}$ as the correlation matrix for patient $n$, where $P$ is the number of regions given by the parcellation. As seen in Fig.~\ref{fig:Fig1}, we model $\mathbf{\Gamma}_{n}$ using a group average basis representation and a patient-specific network strength term. The matrix $\mathbf{B} \in \mathcal{R}^{P \times K}$ is a concatenation of  $K$ elemental bases vectors $\mathbf{b}_{k} \in \mathcal{R}^{P \times 1}$, i.e. $\mathbf{B} := [\mathbf{b}_{1} \quad \mathbf{b}_{2} \quad ... \quad \mathbf{b}_{K}]$, where $K \ll P$. These bases capture steady state patterns of co-activation across regions in the brain. While the bases are common to all patients in the cohort, the combination of these subnetworks is unique to each patient and is captured by the non-negative coefficients $\mathbf{c}_{nk}$. We include a non-negativity constraint $\mathbf{c}_{nk}\geq 0$ on the coefficients to preserve the positive semi-definite structure of the correlation matrices $\{\mathbf{\Gamma}_{n}\}$. Our complete rs-fMRI data representation is:
\begin{equation}
    \mathbf{\Gamma}_{n} \approx \sum_{k}{\mathbf{c}_{nk}\mathbf{b}_{k}\mathbf{b}_{k}^{T}} \ \ \  s.t.  \ \ \ \mathbf{c}_{nk}\geq 0
    \label{eqn1:Eqn1}
\end{equation}
As seen in Eq.~(\ref{eqn1:Eqn1}), we model the heterogeneity in the cohort using a patient specific term in the form of $\mathbf{c}_{n}: = [\mathbf{c}_{n1}\quad ... \quad \mathbf{c}_{nK}]^{T} \in \mathcal{R}^{K \times 1}$. Taking $\textbf{diag}(\mathbf{c}_{n})$ to be a diagonal matrix with the $K$ patient coefficients on the diagonal and off-diagonal terms set to zero, Eq.~(\ref{eqn1:Eqn1}) can be re-written in matrix form as follows:
\begin{equation}
    \mathbf{\Gamma}_{n} \approx {\mathbf{B}\textbf{diag}({\mathbf{c}}_{n})\mathbf{B}^{T}}
    \ \ \  s.t.  \ \ \ \mathbf{c}_{nk}\geq 0
    \label{eqn2:Eqn2}
\end{equation}
Overall, this formulation strategically reduces the high dimensionality of the data, while providing a patient level description of the correlation matrices.

\paragraph{\textbf{Modeling Behavioral Scores}} As shown in the green box of Fig.~\ref{fig:Fig1}, the patient coefficients $\{\mathbf{c}_{nk}\}$ from the representation term, are used to model the clinical severity score $\mathbf{y}_{n}$ using a linear regression vector $\mathbf{w} \in \mathcal{R}^{K \times 1}$
\begin{equation}
\mathbf{y}_{n}\approx{\mathbf{c}_{n}^{T}}\mathbf{w}
\label{eqn3:Eqn3}
\end{equation}
\par Concatenating the vectors $\mathbf{c}_{n}$ into a matrix $\mathbf{C} := [\mathbf{c}_{1} \quad ... \quad \mathbf{c}_{N}] \in \mathcal{R}^{K \times N}$, and the severity scores into a vector $\mathbf{y} := [\mathbf{y}_{1}\quad ... \quad\mathbf{y}_{N}]^{T} \in \mathcal{R}^{N \times 1}$, Eq.~(\ref{eqn3:Eqn3}) can be equivalently represented in matrix form:
\begin{equation}
    \mathbf{y} \approx {\mathbf{C}^{T}}\mathbf{w}
    \label{eqn4:Eqn4}
\end{equation}
\paragraph{\textbf{Joint Objective for Representation and Prediction}}
We combine the two contrasting viewpoints described above into a joint optimization function by summing the contributions of Eq.~(\ref{eqn2:Eqn2}) and Eq.~(\ref{eqn4:Eqn4}) below:
\begin{multline}
\mathcal{J}(\mathbf{B},\mathbf{C},\mathbf{w}) = \sum_{n}{{ {\vert\vert{\mathbf{\Gamma}_{n} - \mathbf{B} \mathbf{diag}(\mathbf{c}_{n})}\mathbf{B}^{T} }\vert\vert}_{F}^{2} } \\ + \gamma{{\vert\vert{\mathbf{y} - \mathbf{C}^{T}\mathbf{w}}\vert\vert}}^{2}_{2} \ \ s.t. \ \ \mathbf{c}_{nk} \geq 0 ,
\label{eqn5:Eqn5}
\end{multline}
Here, $\sum_{n}{\vert\vert{\mathbf{\Gamma}_{n} - {\mathbf{B}\textbf{diag}({\mathbf{c}}_{n})\mathbf{B}^{T}}}\vert\vert}^{2}_{F}$ is the total error in the representation of the $N$ patient correlation matrices, and
${\vert\vert{\mathbf{y} - {\mathbf{C}^{T}}\mathbf{w}}\vert\vert}^{2}_{2}$ is the prediction error for the behavioral data. Finally, $\gamma$ is the trade-off between the rs-fMRI data-representation and score prediction terms. 

\paragraph{\textbf{Regularization Penalties}} Since we wish to capture a compact, yet clinically informative subnetwork representations, we add an $\ell_{1}$ penalty to encourage sparsity in $\mathbf{B}$. Intuitively, this regularizer will sub-select a small number of nonzero entries in $\mathbf{B}$ that explain the data. From an optimization perspective, notice that scaled solution pairs $\{\mathbf{B},\mathbf{C}\}$ and $\{\alpha \mathbf{B}, \frac{1}{\alpha^2} \mathbf{C}\}$, as well as $\{\mathbf{C},\mathbf{w}\}$ and $\{\beta \mathbf{C}, \frac{1}{\beta} \mathbf{w}\}$ give rise to equivalent data representations. As a result, we introduce a quadratic penalty on $\mathbf{C}$ to act as a bound constraint. Similarly, we add an $\ell_{2}$ regularization term to the regression vector $\mathbf{w}$ analogous to ridge regression. Mathematically, the three regularizers can be written as:
\begin{equation}
\lambda_{1}{\vert\vert{\mathbf{B}}\vert\vert}_{1} + \lambda_{2}{\vert\vert{\mathbf{C}}\vert\vert}^{2}_{F} + \lambda_{3}{\vert\vert{\mathbf{w}}\vert\vert}^{2}_{2}
\label{eqn6:Eqn6}
\end{equation}
The penalty terms in Eq.~(\ref{eqn6:Eqn6}) are added to the main objective in Eq.~(\ref{eqn5:Eqn5}). The final joint objective is as follows:
\begin{multline}
    \mathcal{J}(\mathbf{B},\mathbf{C},\mathbf{w}) = \sum_{n}{{ {\vert\vert{\mathbf{\Gamma}_{n} - \mathbf{B} \mathbf{diag}(\mathbf{c}_{n})}\mathbf{B}^{T} }\vert\vert}_{F}^{2} } \\ + \gamma{{\vert\vert{\mathbf{y} - \mathbf{C}^{T}\mathbf{w}}\vert\vert}}^{2}_{2}
    + \lambda_{1}{\vert\vert{\mathbf{B}}\vert\vert}_{1}  \\ + \lambda_{2}{\vert\vert{\mathbf{C}}\vert\vert}^{2}_{F} + \lambda_{3}{\vert\vert{\mathbf{w}}\vert\vert}^{2}_{2} \ \ \ s.t. \ \ \mathbf{c}_{nk} \geq 0 ,
\label{joint_objective}
\end{multline}
The parameter $\lambda_{1}$ controls the number of nonzero elements in $\mathbf{B}$ by scaling the contribution of the $\ell_{1}$ penalty. Similarly, $\lambda_{2}$ and $\lambda_{3}$ relate to element wise bounds on the entries in $\mathbf{C}$ and $\mathbf{w}$ since they scale the contribution of their respective $\ell_{2}$ norms. 
\subsection{{Optimization Algorithm}}
\label{Optim}
\begin{figure*}[t!]
   \centering
   {\includegraphics[width= \textwidth- 5cm]{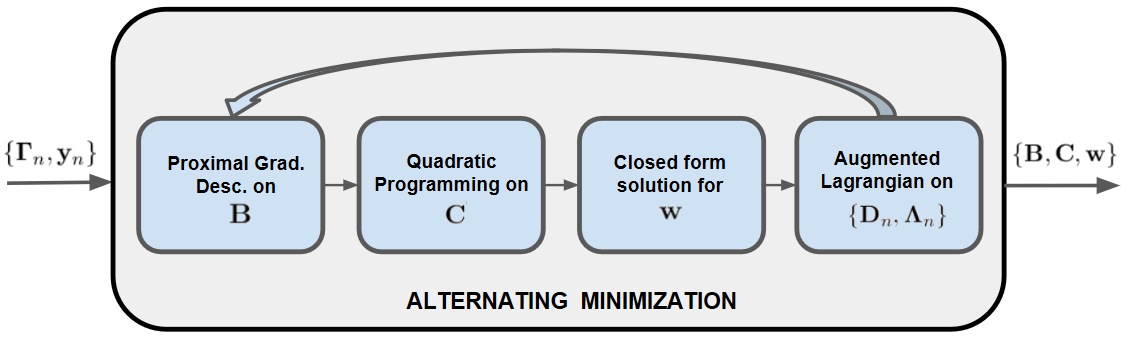}}
   \caption{Our Optimization Strategy, we iterate through four main steps until global convergence}\label{fig2:Fig2}
\end{figure*}
We employ an alternating minimization technique in order to infer the set of latent variables $\{\mathbf{B},\mathbf{C},\mathbf{w}\}$. Here, we optimize the JNO objective function from Eq.~(\ref{joint_objective}) for each output variable, while holding the estimates of the other unknowns constant.
\par Proximal gradient descent [\cite{parikh2014proximal}] is an attractive algorithm to handle the non-differentiable sparsity penalty on ~$\mathbf{B}$ in Eq.~(\ref{joint_objective}), when the supporting terms in the variable of interest are convex. However, from Eq.~(\ref{joint_objective}), we see that the Frobenius norm terms expand to a biquadratic  representation in $\mathbf{B}$, which is non-convex. We circumvent this problem by introducing $N$ constraints of the form $\mathbf{D}_{n} = \mathbf{B}\textbf{diag}(\mathbf{c}_{n})$. We enforce these constraints using the Augmented Lagrangian [\cite{bertsekas2015convex}], denoting the set of Lagrangian matrices by $\{\mathbf{\Lambda}_{n}\}$. The modified objective function in Eq.~(\ref{joint_objective}) takes the form:
\begin{multline}
\mathcal{J}(\mathbf{B}, \mathbf{C},\mathbf{w},\mathbf{D}_{n},\mathbf{\Lambda}_{n}) = {\sum_{n}}{\vert\vert{\mathbf{\Gamma}_{n}-\mathbf{D}_{n}\mathbf{B}^{T}}\vert\vert}_{F}^{2} \\ + \gamma{\vert\vert{\mathbf{y}-\mathbf{C}^{T}\mathbf{w}}\vert\vert}_{2}^{2} + \sum_{n}{\Tr{\left[{\mathbf{\Lambda}_{n}^{T}({\mathbf{D}_{n}-\mathbf{B}\mathbf{diag}(\mathbf{c}_{n})})}\right]}} \\ + \sum_{n}{{\frac{1}{2}}{\vert\vert{\mathbf{D}_{n}-\mathbf{B}\mathbf{diag}(\mathbf{c}_{n})}\vert\vert}_{F}^{2}} + \lambda_{1}{\vert\vert{\mathbf{B}}\vert\vert}_{1}  \\ + \lambda_{2}{\vert\vert{\mathbf{C}}\vert\vert}^{2}_{F} + \lambda_{3}{\vert\vert{\mathbf{w}}\vert\vert}^{2}_{2} \ \ \ \ \  s.t. \ \  \mathbf{c}_{nk} \geq 0
\label{eqn7:Eqn7}
\end{multline}
\par Such that $\Tr[{\mathbf{M}}]$ is the trace operator, which sums the diagonal elements of the argument matrix $\mathbf{M}$. The additional Frobenius norm terms ${\vert\vert{\mathbf{D}_{n}-\mathbf{B}\mathbf{diag}(\mathbf{c}_{n})}\vert\vert}_{F}^{2}$ act as regularizers for the trace constraints. Observe that Eq. (\ref{eqn7:Eqn7}) is now convex in both $\mathbf{B}$ and the set $\{\mathbf{D}_{n}\}$, which allows us to optimize them via standard procedures.
\par Fig.~\ref{fig2:Fig2} provides an overview of the alternating minimization strategy employed. Each individual block in our optimization is described below. We refer the interested reader to Appendix A, which systematically delineates the supporting calculations from this section. 
\paragraph{\textbf{Proximal Gradient Descent on $\mathbf{B}$}}
Given the fixed learning rate parameter $t$, the proximal update for $\mathbf{B}$ is:
\begin{multline}
\mathbf{B}^{k+1} = \mathbf{sgn}(\mathbf{L}).^{*}(\mathbf{max}(\vert{\mathbf{L}}\vert-t,\mathbf{0})) \\  \ \ \ \ s.t.\ \ \ \ \ \ \mathbf{L} = \mathbf{B}^{k} - (t/\lambda_{1})\frac{\partial \mathcal{J}}{\partial \mathbf{B}} \ \ \ \ \ \ \ \ \ \ \ \ \ \ \ \ \ \ 
\label{ProxUpd}
\end{multline}
Here, $-\frac{\partial \mathcal{J}}{\partial \mathbf{B}}$ is a descent direction for the $\mathbf{B}$ update, and $t$ controls the magnitude of the step we take in this direction. In practice, we fix $t$ at $10^{-4}$ for stable convergence. The derivative of $\mathcal{J}$ with respect to $\mathbf{B}$, is computed as:
\begin{multline*}
\frac{\partial \mathcal{J}}{\partial \mathbf{B}} = \sum_{n}\left[{2\left[{\mathbf{B}\mathbf{D}_{n}^{T}\mathbf{D}_{n}-\mathbf{\Gamma}_{n}\mathbf{D}_{n}}\right]-\mathbf{D}_{n}\textbf{diag}(\mathbf{c}_{n})}\right]
\\ +\sum_{n}{\left[\mathbf{B}\textbf{diag}(\mathbf{c}_{n})^{2}-\mathbf{\Lambda}_{n}\textbf{diag}(\mathbf{c}_{n})\right]}
\end{multline*}
At a high level, Eq.~(\ref{ProxUpd}) performs an iterative shrinkage thresholding operation to handle the non-smoothness of the ${\vert\vert{\mathbf{B}}\vert\vert}_{1}$ using a locally smooth quadratic model.
\paragraph{\textbf{Optimizing C using Quadratic Programming}}
The objective is quadratic in $\mathbf{C}$ when $\mathbf{B}$ and $\mathbf{w}$ are held constant. Furthermore, the $\mathbf{diag}(\mathbf{c}_{n})$ term decouples the updates for $\mathbf{c}_{n}$ across patients. We use $N$ quadratic solvers of the form given below to estimate the vectors $\{\mathbf{c}_{n}\}$ : 
\begin{equation}
\frac{1}{2}{\mathbf{c}_{n}^{T}\mathbf{H}_{n}\mathbf{c}_{n}} + \mathbf{f}_{n}^{T}\mathbf{c}_{n} \ \ s.t. \ \ \mathbf{A}_{n}\mathbf{c}_{n} \leq \mathbf{b}_{n}
\label{eqn10:Eqn10}
\end{equation}
The objective and constraint matrices for our quadratic programming solvers are given by:
\begin{eqnarray*}
\mathbf{H}_{n} = \mathcal{I}_{K} \circ (\mathbf{B}^{T}\mathbf{B}) + 2\gamma{\mathbf{w}\mathbf{w}^{T}}+ 2\lambda_{2}\mathcal{I}_{K} \ \ \ \ \     \\  \ \ \ \ \mathbf{f}_{n} = -2\left[\mathcal{I}_{K}\circ(\mathbf{D}_{n}^{T}+\mathbf{\Lambda}_{n}^{T})\mathbf{B}\right]\mathbf{1}  -2\gamma y_n\mathbf{w}; \ \  \\ \mathbf{A}_{n} = -\mathcal{I}_{K} \ \ \ \mathbf{b}_{n} = \mathbf{0}  \ \ \ \ \ \ \ \ \ \ \ \ \ \ 
\label{eqn11:Eqn11}
\end{eqnarray*}
Here, we use $\circ$ to denote the Hadamard product between two matrices and $\mathbf{1}$ to denote a vector of all ones. This strategy helps us find the globally optimal solutions for $\mathbf{c}_{n}$. The constraint matrices $\mathbf{A}_{n}$ and $\mathbf{b}_{n}$ project the solutions onto the $K$ dimensional space of positive reals.
\paragraph{\textbf{Closed Form Update for $\mathbf{w}$}} The global minimizer of $\mathbf{w}$ can be computed by setting the gradient of Eq.~(\ref{eqn7:Eqn7}) equal to zero. Thus, the closed form solution for $\mathbf{w}$ is given by: 
\begin{equation}
\mathbf{w} = (\mathbf{C}\mathbf{C}^{T}+ \frac{\lambda_{3}}{\gamma} \mathcal{I}_{K})^{-1}(\mathbf{C} \mathbf{y})
\end{equation}
This is analogous to a \textbf{regularized linear regression , i.e. ridge regression} update for $\mathbf{w}$.
\paragraph{\textbf{Optimizing the Constraint Variables $\mathbf{D}_{n}$ and $\mathbf{\Lambda}_{n}$}}
Similar to the case of $\mathbf{w}$, each of the primal variables $\{\mathbf{D}_{n}\}$ has a closed form solution given by: 
\begin{equation}
\mathbf{D}_{n} = (\mathbf{diag}(\mathbf{c}_{n})\mathbf{B}^{T}+ 2\mathbf{\Gamma}_{n}\mathbf{B} - \mathbf{\Lambda}_{n})(\mathcal{I}_{K}+2\mathbf{B}^{T}\mathbf{B})^{-1}
\end{equation}
We update the dual variables $\{\mathbf{\Lambda}_{n}\}$ via gradient ascent: 
\begin{equation}
\mathbf{\Lambda}_{n}^{k+1} = \mathbf{\Lambda}_{n}^{k} + \eta_{k}(\mathbf{D}_{n}-\mathbf{B}\mathbf{diag}(\mathbf{c}_{n}))
\end{equation}
The updates for $\mathbf{D}_{n}$ and $\mathbf{\Lambda}_{n}$ ensure that the proximal constraints are satisfied with increasing certainty at each iteration. The learning rate parameter $\eta_{k}$ for the gradient ascent step  of the augmented Lagrangian is chosen to guarantee sufficient decrease for every iteration of alternating minimization. In practice, we initialize this value to $10^{-3}$, and scale it by $0.5$ at each iteration.
\subsection{Prediction on an unseen patient:}
\label{pred_unseen}
\begin{figure*}[t!]
   \centering
   \fbox{\includegraphics[width= \textwidth- 5cm]{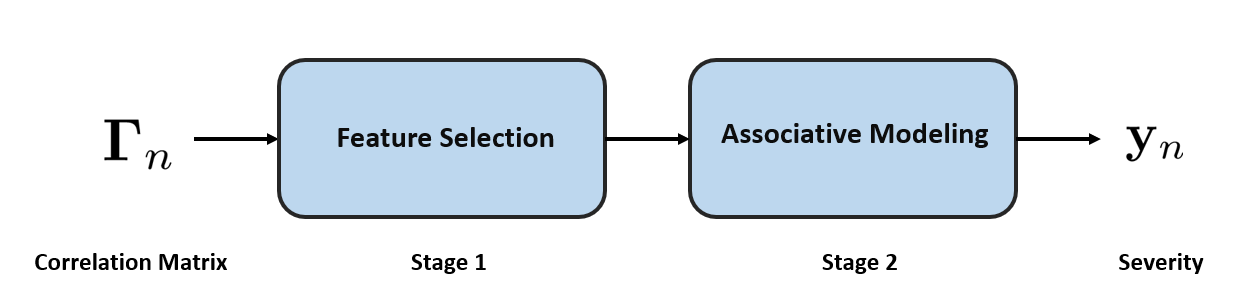}}
   \caption{A typical two stage baseline. We input the correlation matrices to Stage~$1$, which performs Feature Extraction on the raw correlations. This step could be a technique from machine learning, graph theory or a statistical measure. Stage~$2$ fits an associative regression model to the output representation of Stage~$1$ }\label{BS}
\end{figure*}
In order to estimate the coefficients ${\hat{\mathbf{c}}}$ for a new patient, we re-solve the quadratic program in Eq.~(\ref{eqn10:Eqn10}) using the $\{\mathbf{B}^{*},\mathbf{w}^{*}\}$ computed from the training data via the procedure outlined in Section \ref{Optim}. We explicitly set the contribution from the data term in Eq.~(\ref{eqn5:Eqn5}) to $0$, since the corresponding value of $\mathbf{\hat{y}}$ is unknown for the new patient. We also implicitly assume that the conditions for the proximal operator hold, i.e. the constraint $\mathbf{\hat{D}} = \mathbf{B}^{*}\mathbf{diag}(\mathbf{\hat{c}})$ is exactly satisfied. The estimation of the unseen patient's coefficients are thus mathematically formulated as follows:
\begin{multline}
\mathbf{\hat{c}} = \argmin_{\mathbf{c}} {{ {\vert\vert{\mathbf{\Gamma}_{n} -  \mathbf{B} \mathbf{diag}(\mathbf{c})}\mathbf{B}^{T} }\vert\vert}_{F}^{2}} + \lambda_{2}{\vert\vert{\mathbf{c}}\vert\vert}^{2}_{2} \\ \ \ \ s.t. \ \ \mathbf{c}_{k} \geq 0 \ \ \ \ \ \ \ \
\label{ctest}
\end{multline}
Once again, Eq.~(\ref{ctest}) can be formulated as a quadratic program. The parameters from Eq.~(\ref{eqn10:Eqn10}) correspond to:
\begin{eqnarray*}
\ \ \ \ \ \ \ \ \ \ \ \ \mathbf{H}_{n} = 2(\mathbf{B}^{T}\mathbf{B})\circ(\mathbf{B}^{T}\mathbf{B}) + 2\lambda_{2}\mathcal{I}_{K} \ \ \ \ \ \   \\ \mathbf{f}_{n} = -2\left[\mathcal{I}_{K}\circ(\mathbf{B}^{T}\mathbf{\Gamma}_{n}\mathbf{B})\right]\mathbf{1}; \ \ \ \ \ \ \ \ \  \ \ \  \\  \ \ \ \ \ \ \ \  \mathbf{A}_{n} = -\mathcal{I}_{K} \ \ \ \ \ \ \  \mathbf{b}_{n} = \mathbf{0} \ \ \ \ \ \ \ \ \ \ \ \ \ \   
\end{eqnarray*}
The estimate for the behavioral score for the test patient is given by the vector product $\mathbf{\hat{y}}= \mathbf{\hat{c}}^{T}\mathbf{w}^{*}$.
\subsection{Baseline Comparison Techniques}
\label{baselines}
We evaluate the performance of our method against a set of well established statistical, graph theoretic, and data-driven frameworks that have been used to provide rich feature representations. Fig.~\ref{BS} describes a general two stage pipeline for our task. The first stage is a representation learning step used for feature extraction. Stage~$2$ is a regression model to map the learned features to behavioral data. We evaluate our method against several choices of linear and non-linear algorithms for Stage~$1$. These are combined with a regularized linear regression in Stage~$2$, similar to our method. Additionally, we evaluate the performance obtained by omitting a Stage~$1$ and training a deep neural network end-to-end on the input correlation features. Lastly, we demonstrate the advantage provided by combining the neuroimaging and behavioral representations in the JNO framework. For this, we present a comparison where the feature learning and prediction stages are decoupled, similar to the baselines.
\subsubsection{Machine Learning Approach (PCA):}
\label{ML}
We start with the ${P \times P}$ correlation matrix $ \mathbf{\Gamma}_{n}$ for each patient. Since this matrix is symmetric, we have $M =\frac{P \times (P-1)}{2}$ distinct rs-fMRI correlation pairs between various communicating sub-regions. Accordingly, the features from every individual are composed into a descriptor matrix $\mathbf{X} \in \mathcal{R}^{M \times N}$. We further concentrate these feature into a small number of representative bases. The basis extraction procedure in Stage~$1$ corresponds to a linear mapping in the original correlation space via a \textbf{Principal Component Analysis (PCA)}. In Stage~$2$, we construct a \textbf{regularized linear regression (ridge regression)} on the projected features to predict the clinical severity. PCA projects the observations onto a set of uncorrelated \textit{principal component basis} by means of an orthogonal linear decomposition. Mathematically, PCA poses the following dimensionality reduction problem:
\begin{multline}
\mathcal{F}(\mathbf{U},\mathbf{Z},\boldsymbol{\mu}) = \argmin_{\boldsymbol{\mu},\mathbf{U},\mathbf{Y}} {{\vert\vert{\mathbf{X}-\boldsymbol{\mu}\mathbf{1}^{T}-\mathbf{U}\mathbf{Z}}\vert\vert}^{2}_{F}} \\ \ \ \  s.t. \ \  \mathbf{U}^{T}\mathbf{U} = \mathcal{I}_{d}, \ \  \mathbf{Z}\mathbf{1} =\mathbf{0}
\label{eqn12:Eqn12}
\end{multline}
Here, $\mathbf{U} \in \mathcal{R}^{N \times d}$ is the $d$ dimensional subspace basis which best approximates the information from $\mathbf{X}$ in the Frobenius norm sense, computed by calculating the eigenvectors of the sample covariance matrix $\mathbf{X}\mathbf{X}^T$. Consequently, $\mathbf{Z} \in \mathcal{R}^{d \times N}$ is a compact $d$ dimensional representation of $\mathbf{X}$, where $d \ll M$. $\mathbf{1}$ is a $d$ dimensional vector of ones. The constraint $\mathbf{Z}\mathbf{1} =\mathbf{0}$ centers the representation $\mathbf{Z}$.
\subsubsection{Statistical Approach (ICA)}
Here, we use \textbf{Independent Component Analysis (ICA)} as the Stage~$1$ algorithm combined with \textbf{ridge regression}. ICA operates on the raw time series data to extract representative spatial patterns that explain rs-fMRI connectivity. ICA has become ubiquitous for identifying group level as well as individual-specific connectivity signatures. It decomposes a multivariate signal into `independent' non-Gaussian components based on the statistics of the data. Mathematically, ICA models the components $\{\mathbf{y}_{k}\}$ of the observed signal $\mathbf{y} = \big[ \mathbf{y}_{1}, \dots ,\mathbf{y}_{m} \big]$ as a sum of $n$ independent components $\mathbf{S} = \big[\mathbf{s}_{1}, \dots ,\mathbf{s}_{n} \big]$ combined via the mixing matrix $\mathbf{A} = \big[ \mathbf{a}_{1}, \dots ,\mathbf{a}_{n}\big]$
\begin{equation}
\mathbf{y} = \sum_{i=1}^{n}{\mathbf{s}_{i}\mathbf{a}_{i}}
\ \ \ \  \textit{i.e.} \ \  \mathbf{Y} = \mathbf{A}\mathbf{S}
\label{ICA}    
\end{equation}
$\mathbf{s}$ can be recovered by multiplying the observed signals $\mathbf{Y}$ with the inverse of the mixing matrix $\mathbf{W} =\mathbf{A}^{-1}$. We adaptively estimate both the mixing matrix $\mathbf{A}$ and the components $\mathbf{s}$ by setting up a cost function that maximizes the non-gaussianity of $\mathbf{s}_{i} = \mathbf{w}_{i}^{T}\mathbf{y}$ or minimizes the mutual information.
\par Group ICA extends this algorithm to a multi-subject analysis for extracting independent spatial patterns common across patients, but combined via individual time courses. We use the GIFT [\cite{calhoun2009review}] software in order to perform Group-ICA to derive independent spatial maps for each patient. The correlation values between the identified components are fed to the regression model.

\subsubsection{Graph Theoretic Approach (Node Degree):}
\label{GT}
Each correlation matrix $\mathbf{\Gamma}_{n}$ can be thresholded and considered a graph adjacency matrix, which we denote by $\mathbf{\Psi} \in \mathcal{R}^{P \times P}$. The element $\mathbf{\Psi}_{ij}$ gives the strength of association between two communicating sub-regions $i$ and $j$. The underlying graph topology can be summarized using node/edge based importance measures [\cite{sporns2004organization} \cite{bassett2006small}]. Again, we use a regularized linear regression technique to estimate the severity score from the reduced representation. This treatment closely parallels the machine learning approach, as we can view the graph measures as a dimensionality reduction. We compute \textbf{Node Degree ($D_{N}$)} from the adjacency graph followed by a \textbf{ridge regression} on the features.
\par Given the adjacency matrix $\mathbf{\Psi}$, the degree of region $v$ is equal to the number of edges incident on $v$, with loops counted twice. Mathematically, the degree $\mathbf{D}_{N}(v)$ is computed as follows:
\begin{equation}
\mathbf{D}_{N}(v) = \sum_{j \neq v}{\mathbb{1}(\mathbf{\Psi}_{jv}>0)}
\end{equation}
where, $\mathbb{1}(.)$ is the indicator function, which takes the value $1$ if the condition is satified, and $0$ otherwise. This metric captures the importance of each node in explaining the graph, which in our case, corresponds to the average connectivity strength of each region in the brain.

\subsubsection{A Neural Network Approach:}
\label{ANN}
Recently, there has been an upsurge in using neural networks to investigate neuroimaging correlates of developmental disorders [\cite{kaiser2010neural}]. Here, we test the efficacy of a simple \textbf{Artificial Neural Network (ANN)} for predicting the severity score from the correlation feature matrix $\mathbf{X}$ defined above. The network architecture encodes a series of non-linear transformations of the input correlations to approximate the severity score. Recall that the size of the input is dependent on our choice of parcellation, which could be of considerable width (of the order of $\approx{5000}$ connections for $P=100$). After evaluating several architectures, we employ a two hidden layer network with widths $8000$ and $10$ respectively. We use a Rectified Linear Unit (ReLU) non-linearity after the first hidden layer and a Tanh non-linearity after the second hidden layer. We used the ADAM optimizer with an initial learning rate of $10^{-4}$, scaled by $0.9$ per $10$ epochs, and a momentum of $0.9$ to train the network.

\section{Experiments:}
\subsection{Validation on Synthetic Data}
\begin{figure}[b!]
   \centering
   \includegraphics[scale=0.6]{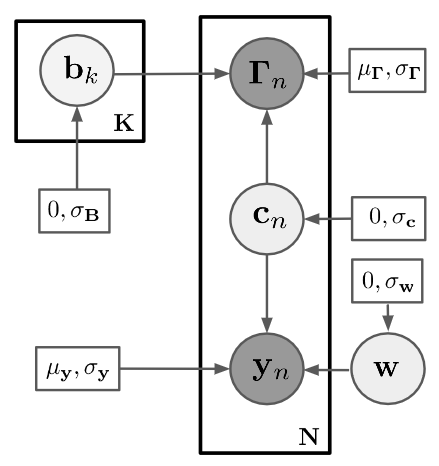}
   \caption{The graphical model for the joint objective. For our synthetic experiments, we fix the model parameters $\mathbf{\sigma}_{\mathbf{C}}= 2, \mathbf{\sigma}_{\mathbf{w}}= 0.2$}\label{fig3:Fig3}
\end{figure}
As a sanity check, we first sample data from the generative model in Eq.~(\ref{joint_objective}) and use the optimization outlined in Section~\ref{Optim} to estimate the unknowns $\{\mathbf{B},\mathbf{C},\mathbf{w}\}$. This procedure helps us analyze the performance of the algorithm under different noise scenarios. The inputs to our model are the correlation matrices $\{\mathbf{\Gamma}_{n}\}$ and the clinical scores $\{\mathbf{y}_{n}\}$. We note that the model gives a complete description of each $\mathbf{\Gamma}_{n}$ in terms of the basis vectors $\{\mathbf{b}_{k}\}$ and the patient coefficients $\{\mathbf{c}_{n}\}$. Since the data representation terms for each patient are coupled solely through the basis representation, the coefficient descriptors are independent of each other. In a similar observation, each score $\mathbf{y}_{n}$ is explained by the corresponding $\mathbf{c}_{n}$, independent of the remaining subjects, when we fix the regression vector $\mathbf{w}$. We use this information to describe the \textit{observed data} $\{\mathbf{\Gamma}_{n},\mathbf{y}_{n}\}$ using a generative model with the likelihood model based on the \textit{hidden variables} $\{\mathbf{B},\mathbf{C},\mathbf{w}\}$.
\begin{figure*}[t!]
   \centering
   \fbox{\includegraphics[width= \textwidth]{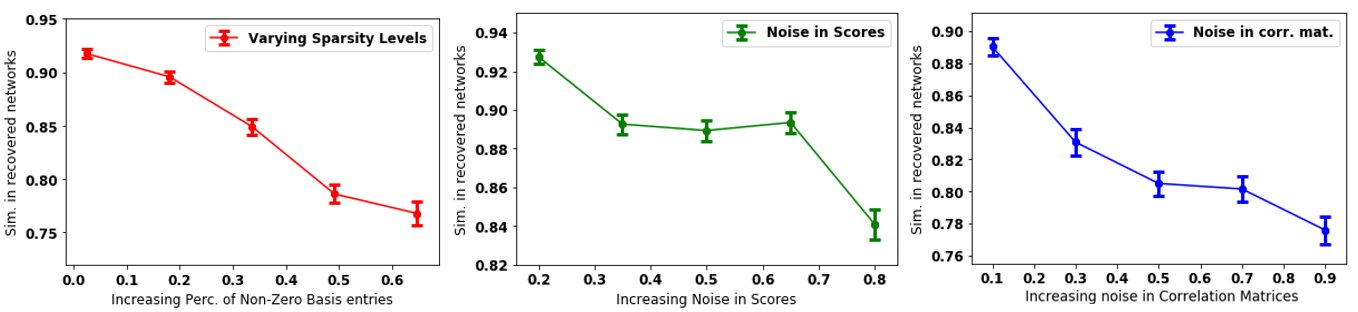}}
   \caption{Performance on synthetic experiments. (\textbf{L}): Varying the level of sparsity  ($\mathbf{\sigma}_{\mathbf{\Gamma}_{n}}=0.4$, $\mathbf{\sigma}_{\mathbf{y}_{n}} = 0.2$), (\textbf{M}): Varying the level of noise in $\mathbf{y}_{n}$ ($\mathbf{\sigma}_{\mathbf{B}} =0.2$, $\mathbf{\sigma}_{\mathbf{\Gamma}_{n}}=0.4$) , (\textbf{R}): Varying the level of noise in $\mathbf{\Gamma}_{n}$ under ($\mathbf{\sigma}_{\mathbf{B}} =0.2$, $\mathbf{\sigma}_{\mathbf{y}_{n}} = 0.2$) Values on the x-axis have been normalized to reflect a $[0-1]$ range by dividing by the maximum value of the variable. Deviations from the mean recovered similarity for each parameter setting is indicated in the figure and have been reported as a standard error value. The reported $x$-axis range reflects the regimes within which the algorithm converges to a local solution }
   \label{fig4:Fig4}
\end{figure*}
\par Notice that, when treated as a Bayesian log-likelihood (i.e. taking a negative exponent of the objective), the $\ell_{2}$ norms in Eq.~(\ref{joint_objective}) translate into Gaussian distributions, and the $\ell_{1}$ norm is equivalent to a Laplacian prior. The corresponding graphical model is shown in Fig.~\ref{fig3:Fig3}. The observed variables are indicated by the shaded circles. The white circles contain the hidden variables. The distribution parameters for the hidden variables are indicated in the corresponding rectangle pointing to the variable. The Laplacian parameter $\sigma_{B}$ controls the overlap in the patterns of sparsity in $\mathbf{B}$, which relates to $\lambda_{1}$. $\mathbf{C}$ and $\mathbf{w}$ are described by Gaussians with means zero (i.e. $\ell_{2}$ norm offset). The variances $\sigma^{2}_{\mathbf{C}}$ and $\sigma^{2}_{\mathbf{w}}$ are related to the penalty parameters $\lambda_{2}$ and $\lambda_{3}$ respectively. The non-negativity constraint on $\mathbf{c}_{n}$ is handled by folding (i.e. taking the absolute value of) the normal distribution to restrict the $\mathbf{c}_{n}$ values to be positive reals. The observed variable $\{\mathbf{y}_{n}\}$, translates to a Gaussian with mean $\mathbf{\mu}_{\mathbf{y}_{n}}=\mathbf{c}_{n}^{T}\mathbf{w}$, and variance parameters $\sigma_{\mathbf{y}_{n}}$. This is again folded to reflect positive values of $\mathbf{y}_{n}$. The correlation matrices $\{\mathbf{\Gamma}_{n}\}$ are drawn from a Gaussian distribution with mean $\mathbf{\mu}_{\mathbf{\Gamma}_{n}}= \mathbf{B}\mathbf{diag}(\mathbf{c}_{n})\mathbf{B}^{T}$ (which is positive semi-definite by construction) and variance $\mathbf{\sigma}_{\Gamma_{n}}$.
\par There are two sources of noise for the observed variables, which include the error in the correlation matrices $\mathbf{\Gamma}_{n}$, and the error in the severity scores $\mathbf{y}_{n}$. These scenarios can be directly related to controlling the variance parameters $\sigma_{\mathbf{\Gamma}_{n}}$ and $\sigma_{\mathbf{y}_{n}}$ respectively. Additionally, we are interested in the performance of the algorithm under varying levels of overlap in the sparsity patterns in $\mathbf{B}$.  
\par We evaluate the performance using an average inner-product measure of similarity $S$ between each recovered network, $\mathbf{\hat{b}}_{k}$, and its corresponding best matched generating network, $\mathbf{b}_{k}$, both normalized to unit norm, i.e.:
\begin{equation}
{S} ={{\frac{1}{K}}\sum_{k}\frac{\vert{ \mathbf{b}_{k}^{T}  \mathbf{\hat{b}}_{k}}\vert}{{\vert\vert{\mathbf{b}_{k}}\vert\vert}_{2}{\vert\vert{\mathbf{\hat{b}}_{k}}\vert\vert}_{2}}}{.}
\label{eqn22:Eqn22}
\end{equation}
Fig.~\ref{fig4:Fig4} depicts the performance of the algorithm in these three cases. The $x$-axis corresponds to increasing the levels of noise, while the $y$-axis indicates the similarity metric $S$ computed for the particular setting. In the leftmost plot, an $x$-axis value close to $0$ indicates high percentage of sparsity in $\mathbf{B}$, while increasing values correspond to denser basis matrices. Throughout this experiment, the values of the other free parameters in the generative model were held constant. The middle plot evaluates subnetwork recovery when the noise in the scores, i.e. $\mathbf{\sigma}_{\mathbf{y}_{n}}$ is increased. The x-axis reports normalised values of $\mathbf{\sigma}_{\mathbf{y}_{n}}$ while the remaining free parameters were held constant. Similarly, the rightmost plot in Fig.~\ref{fig6:Fig6} indicates performance under varying noise in the correlation matrices $\mathbf{\Gamma}_{n}$. Again, normalized $\mathbf{\sigma}_{\mathbf{\Gamma}_{n}}$ values are reported on the x-axis. All numerical results have been aggregated over $100$ independent trials.
\par As expected, increasing the noise in the correlation matrices and scores worsens the recovery performance of the algorithm. This is indicated by the decay in the similarity measure with increasing noise parameters as well as an increase in the corresponding variance. Additionally, the algorithm performs better when there is lesser overlap in the columns of $\mathbf{B}$, i.e. when the generating basis is sparse. However, we observe that our algorithm is robust in the noise regime estimated from the real-world rs-fMRI data $(0.01-0.2)$ and recovered sparsity levels $(0.1-0.4)$. In addition, we identify the stable parameter settings for the algorithm which guide our real world experiments.

\subsection{A Population Study of Autism :}
\par We evaluate the efficacy of our JNO framework on two separate cohorts. Our first dataset consists of $58$ children with high functioning ASD acquired at the Kennedy Krieger Institute in Baltimore, USA. We refer to this as the KKI dataset. The age of the subjects ranged from $10.06 \pm 1.26$ and the IQ as $110\pm14.03$. The second cohort is a subset of the publicly available Autism Brain Imaging Data Exchange (ABIDE I) [\cite{di2014autism}] acquired at the New York University (NYU) site consisting of $63$ patients. Social and communicative deficits in autism are believed to arise from abnormal interactions between functionally linked regions in the brain [\cite{pelphrey2014building}]. Therefore, identifying long-range correlative patterns in the brain directly linked to clinical severity is an important stepping stone to understanding and quantifying the neural underpinnings of the disorder. 
\paragraph{\textbf{Neuroimaging Data}}  
For the KKI dataset, rs-fMRI scans were acquired on a Phillips $3T$ Achieva scanner using a single shot, partially parallel gradient-recalled EPI sequence with TR/TE = $2500/30$ms, flip angle $\ang{70}$, res $=3.05 \times 3.15 \times 3$mm, having $128$ or $156$ time samples. The children were instructed to relax with eyes open and focus on a central cross-hair while remaining still for the duration of the scan.
\par Slice time correction, rigid body realignment, and normalization to the EPI version of the MNI template was performed as a part of pre-processing using SPM [\cite{penny2011statistical}]. The time courses were temporally detrended in order to remove gradual trends in the data. From here, we used a CompCorr [\cite{behzadi2007component}] strategy for the estimation and removal of spatially coherent noise from the white matter and ventricles, along with the linearly detrended versions of the six rigid body realignment parameters and their first derivatives. We performed a spatial smoothing with a $6$mm FWHM Gaussian kernel followed by a temporal filtering using a $0.01-0.1$Hz pass band. Lastly, we removed spikes from the data via tools from the AFNI package [\cite{cox1996afni}] as an alternative to motion scrubbing.
\par For the NYU cohort, the rs-fMRI data was pre-processed using the configurable pipeline for the analysis of connectomes, that has been integrated with ABIDE. The pre-processing steps involved are skull-stripping, global mean intensity normalization, spatial normalization to the MNI template, nuisance regression and CompCorr, followed by bandpass filtering, but without global signal regression. 
\par This work relies on the Automatic Anatomical Labelling (AAL) atlas [\cite{tzourio2002automated}], which defines $116$ cortical, subcortical and cerebellar regions. We compute a $116 \times 116$ correlation matrix for each patient based on the Pearson's Correlation Coefficient between the average time series for these regions. Empirically, we observed a consistent noise component with nearly constant contribution from all brain regions and low predictive power for both datasets. Therefore, we subtracted out the first eigenvector contribution from each of the correlation matrices and used the residuals as the inputs $\{\mathbf{\Gamma}^{t}_{n}\}$ to the algorithm and the baselines.
\paragraph{\textbf{Behavioral Data}} We analyzed three independent measures of clinical severity. These include:
\begin{itemize}
\setlength\itemsep{0.1em}
    \item[1] Autism Diagnostic Observation Schedule, Version $2$ (ADOS-2) total raw score
    \item[2] Social Responsiveness Scale (SRS) total raw score
    \item[3] Praxis total percent correct score
\end{itemize} 
ADOS and SRS are standard assessments and are available for both the KKI and NYU datasets, while the Praxis has been collected for the KKI dataset only. 
\par The ADOS consists of different sub-scores which  quantify the social and communicative deficits of the patient along with the restrictive/repetitive behaviors [\cite{lord2000autism}]. The test is administered by a trained clinician who evaluates the child against a set of guidelines. The total score is computed by adding the individual sub-scores. The dynamic range for ADOS is between $0$ and $30$, with higher score indicating greater impairment.
\par The SRS scale characterizes the social responsiveness of an individual [\cite{bolte2008assessing}]. Typically, a parent/care-giver or teacher completes out a standardized questionnaire assessing various aspects of the child's behavior. SRS reporting tends to be more variable across patients compared to ADOS, since the responses are heavily biased by the parent/teacher attitudes. The SRS dynamic range is between $70-200$ for ASD patients, with higher values corresponding to higher severity in terms of social responsiveness.
\par Praxis was assessed using the Florida Apraxia Battery (modified for children) [\cite{mostofsky2006developmental}], which assesses ability to perform skilled motor gestures to command, to imitation, and with actual tool use. Several studies (\cite{mostofsky2006developmental}, \cite{dziuk2007dyspraxia}, \cite{dowell2009associations}, \cite{nebel2016intrinsic} etc) reveal that children with ASD show marked impairments in Praxis i.e., developmental dyspraxia, and that impaired Praxis correlates with impairments in core autism social-communicative and behavioral features. Performance is videotaped and later scored by two trained research-reliable raters, with total percent correctly performed gestures as the dependent variable of interest. Scores therefore range from $0-100$, with higher scores indicating better Praxis performance. This measure was available for $52$ of the total patients in the KKI dataset. 
\subsection {Evaluating Predictive Performance}
We wish to compare the performance of our JNO framework against a wide class of algorithms described in Section~\ref{baselines}. In this paper, we use regularized linear regression (i.e. ridge regression) for all baselines, except the ANN. Our Supplementary Results document includes further comparison with a non-linear (Random Forest) regression model. We emphasize that the baseline performance is nearly identical for both the linear and the non-linear models.
\begin{figure}[h!]
   \centering
   \fbox{\includegraphics[height = 3.25cm,width =8 cm]{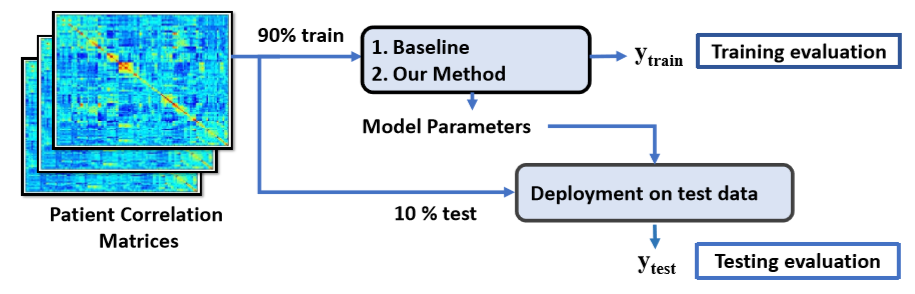}}
   \caption{A ten-fold cross validation for evaluating performance}\label{fig5:Fig5}
\end{figure}
\par We characterize the performance of each method using a $10$ fold cross validation strategy as illustrated in Fig.~\ref{fig5:Fig5}. For a given parameter setting, we first split the data set into $10$ training and test folds. For each of the folds, we train the models on a $90$ percent training set split of the data. We report the score prediction on the held out $10$ percent, which constitutes the testing set for that fold. Note that each datapoint is a part of the test set in exactly one of the 10 folds.
\par We report two quantitative measures of performance. Median Absolute Error (MAE) quantifies the absolute distance between the measured and predicted scores: 
\begin{equation*}
\mathrm{MAE} = \mathrm{median}(\vert{\mathbf{\hat{y}}-\mathbf{y}}\vert),
\end{equation*}
where the median is computed across the set of patients. We report MAE values with the standard deviation of the error. Lower MAE indicates better testing performance.
\par Normalized Mutual Information (NMI) assesses the similarity in the distribution of the predicted and observed score distributions across test patients. NMI is computed as follows:
\begin{equation*}
\mathrm{NMI}(\mathbf{y},\mathbf{\hat{y}}) = \frac{H(\mathbf{y}) + H(\mathbf{\hat{y}}) - H(\mathbf{y},\mathbf{\hat{y}})}{\min{\{H(\mathbf{y})},H(\mathbf{\hat{y}}) \}}  
\end{equation*}
where $H(\mathbf{y})$ denotes the entropy of $\mathbf{y}$ and $H(\mathbf{y},\mathbf{\hat{y}})$ is the joint entropy between $\mathbf{y}$ and $\mathbf{\hat{y}}$. NMI ranges from $0-1$ with higher values indicating a better agreement between predicted and measured score distributions, and thus characterizing improved performance.

\begin{figure}[t]
  \centering
    \includegraphics[width=0.48\textwidth]{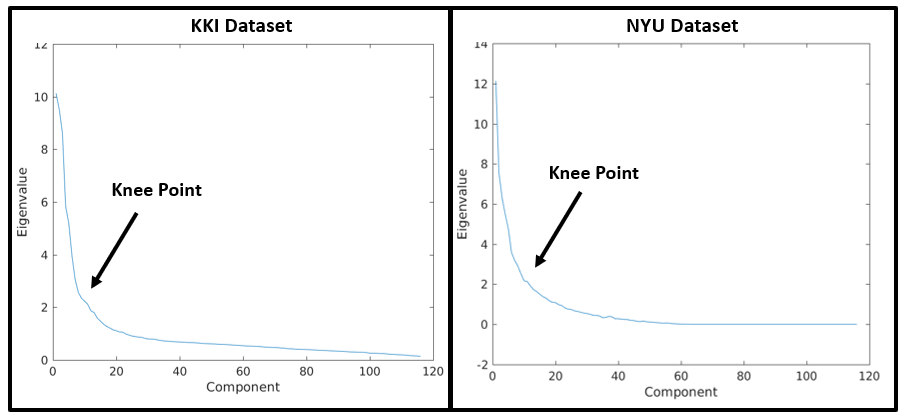}
    \caption{Scree Plot of the correlation matrices to corrobrate the selected values for K. \textbf{(L)} KKI Dataset \textbf{(R)} NYU Dataset}
    \label{ScreePlot}
\end{figure}

\par Along with MAE and NMI, we perform a statistical test on the error distribution to evaluate the performance gain of our framework. We first calculate the Cumulative Distribution Function (CDF) of errors and then report the Kolmogorov-Smirnov statistic on the CDF. This statistic helps us quantify the differences in the error distributions of each baseline compared to our algorithm. A Kolmogorov-Smirnov statistic lower than $\alpha = 0.05$ is widely accepted in literature as statistically significant. We indicate comparisons which fall within this threshold in bold and near misses using an underline.
\begin{figure*}[b]    
    \centering
      \includegraphics[width=\dimexpr \textwidth-2\fboxsep-2\fboxrule\relax, scale=0.8]{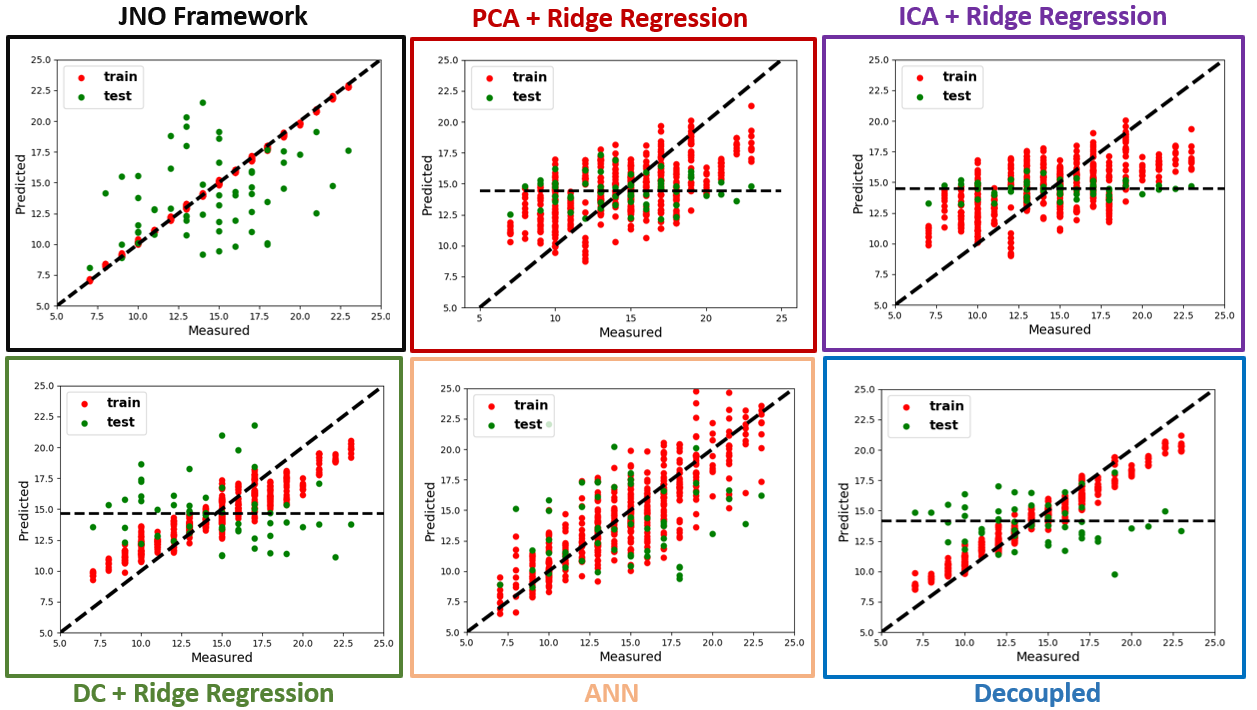}
   \caption{\textbf{KKI dataset:} Prediction performance for the ADOS score for \textbf{Black Box:} JNO Framework. \textbf{Red Box:} PCA and ridge regression \textbf{Purple Box:} ICA and ridge regression \textbf{Green Box:} Node degree centrality and ridge regression \textbf{Orange Box}: ANN on correlation features \textbf{Blue Box:} Decoupled matrix cactorization and ridge Regression} \label{fig6:Fig6}
\end{figure*}
\begin{figure*}     
    \centering
      \includegraphics[width=\dimexpr \textwidth-2\fboxsep-2\fboxrule\relax, scale=0.785]{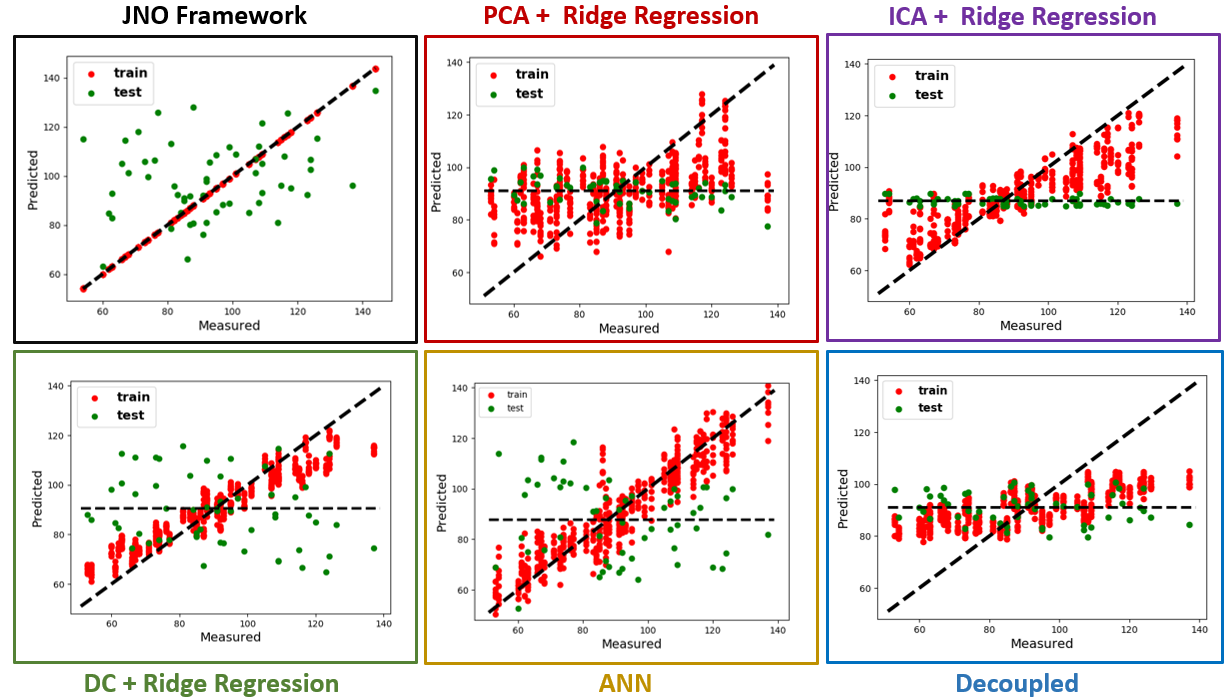}
\caption{\textbf{KKI dataset:} Prediction performance for the SRS score for \textbf{Black Box:} JNO Framework. \textbf{Red Box:} PCA and ridge regression \textbf{Purple Box:} ICA and ridge regression \textbf{Green Box:} Node degree centrality and ridge regression \textbf{Orange Box}: ANN on correlation features \textbf{Blue Box:} Decoupled matrix factorization and ridge regression} \label{fig7:Fig7}
\end{figure*}

\subsubsection{Parameter Settings}

Our method has five user-specified parameters $\{\gamma,\lambda_{1},\lambda_{2},\lambda_{3},K\}$. Recall that $K$ is the number of basis networks, $\gamma$ is the penalty tradeoff between the representation and regression terms, $\lambda_{1}$ is the sparsity penalty, while $\lambda_{2}$ and $\lambda_{3}$ are the regularization penalties on the coefficients $\mathbf{C}$ and regression weights $\mathbf{w}$ respectively.

\par We use the knee point of the eigenspectrum of the correlation matrices $\mathbf{\Gamma}_{n}$ to select the number of bases ($K=8$ for both datasets). For reference, we have included the scree-plots in Fig.~$8$. Empirically, the JNO model is insensitive to the choice of $\lambda_{3}$ and $\gamma$, so we fix both at one. Effectively, we are left with two free parameters, which we optimize by performing a bivariate grid search. We note that the generalization accuracy is dependent on the dynamic range of the scores and is sensitive to $\lambda_{1}$ and $\lambda_{2}$. As described in Section $3.6$, we have identified a stable range of operation across a single order of magnitude for these parameters. Based on the cross validation results, we finally use the following settings in our experiments: For the KKI dataset, $\{{\lambda}_{2}=0.2,{\lambda}_{1}=30\}$ for ADOS, $\{{\lambda}_{2}=0.9,{\lambda}_{1}=50\}$ for SRS and $\{{\lambda}_{2}=0.6,{\lambda}_{1}=20\}$ for Praxis; for the NYU dataset, $\{{\lambda}_{2}=0.1,{\lambda}_{1}=20\}$ for ADOS, $\{{\lambda}_{2}=0.9,{\lambda}_{1}=40\}$ for SRS. 

\par To provide a fair comparison with our JNO framework, we use a joint grid search on the Stage~$1$ hyperparameters and the Stage~$2$ ridge penalty to optimize these values for every baseline method. Again, we report the best performance in a ten fold cross validation setting.

\par We select $10$ PCA components for the KKI dataset, and $15$ for the NYU dataset. For ICA, we obtained good performance for $35$ spatial maps obtained from GIFT [\cite{calhoun2009review}]. For the graph theoretic baseline, we threshold the correlation matrices $\{\mathbf{\Gamma}_{n}\}$ at $0.2$ to obtain valid adjacency matrices $\{\mathbf{\Psi}_{n}\}$. In conjunction with these, the ridge penalty parameter was swept across four orders of magnitude.

\par As described in Section~\ref{ANN}, we fixed the network architecture to be a two hidden layer network with widths $8000$ and $10$ respectively, having a Rectified Linear Unit (ReLU) non-linearity after the first hidden layer and a Tanh non-linearity after the second hidden layer. We use a standard weight decay regularizer, with the regularization parameter varied over three orders of magnitude for each baseline comparison. We trained the network using the ADAM optimizer with an initial learning rate of $10^{-4}$, scaled by $0.9$ per $10$ epochs, and a momentum of $0.9$.

\par Finally, we include the performance upon decoupling the ridge regression and the matrix decomposition in Eq.~(\ref{eqn5:Eqn5}) as a sanity check. This is akin to the two stage treatment in the baselines where the two terms are not explicitly coupled as in the JNO objective.

\subsection{Performance on Real World Data}
\label{Results}

\begin{figure*}    
    \centering
      \includegraphics[width=\dimexpr \textwidth-2\fboxsep-2\fboxrule\relax, scale=0.8]{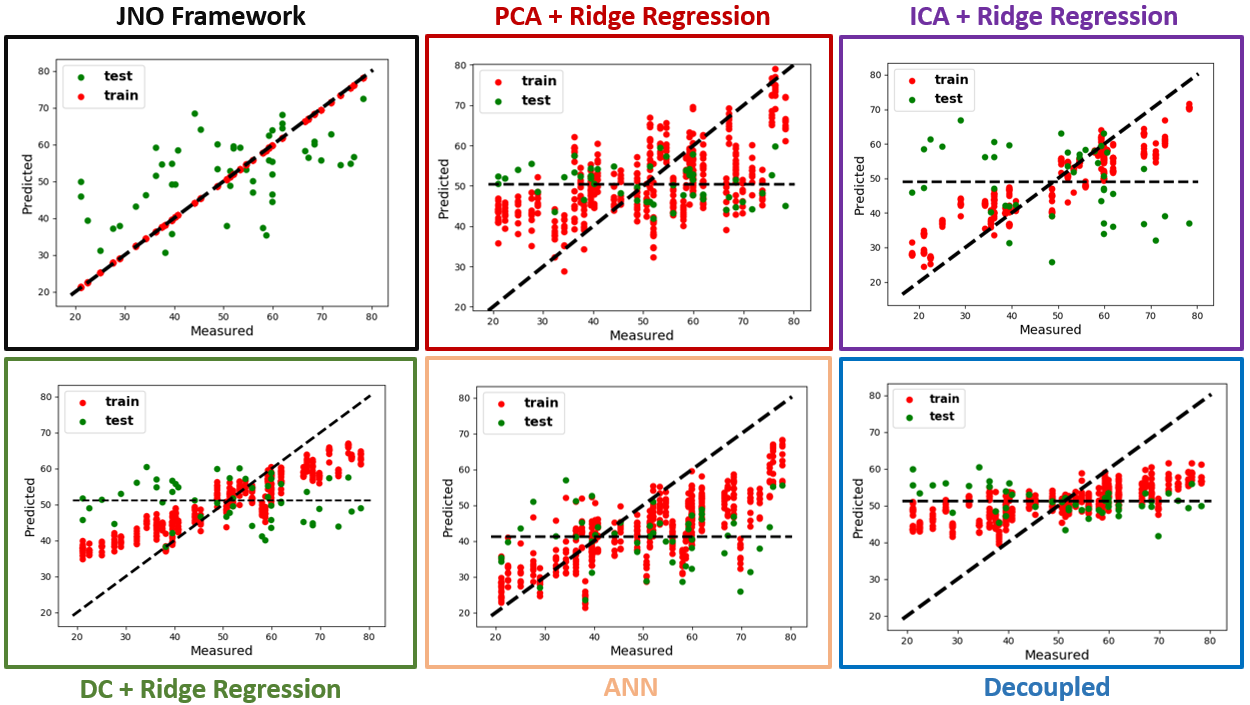}
   \caption{ \textbf{KKI dataset:} Prediction performance for the Praxis score for \textbf{Black Box:} JNO Framework. \textbf{Red Box:} PCA and ridge regression \textbf{Purple Box:} ICA and ridge regression \textbf{Green Box:} Node degree centrality and ridge regression \textbf{Orange Box}: ANN on correlation features \textbf{Blue Box:} Decoupled matrix factorization and ridge regression} \label{fig8:Fig8}
\end{figure*}

\begin{figure*}    
    \centering
      \includegraphics[width=\dimexpr \textwidth-2\fboxsep-2\fboxrule\relax, scale=0.8]{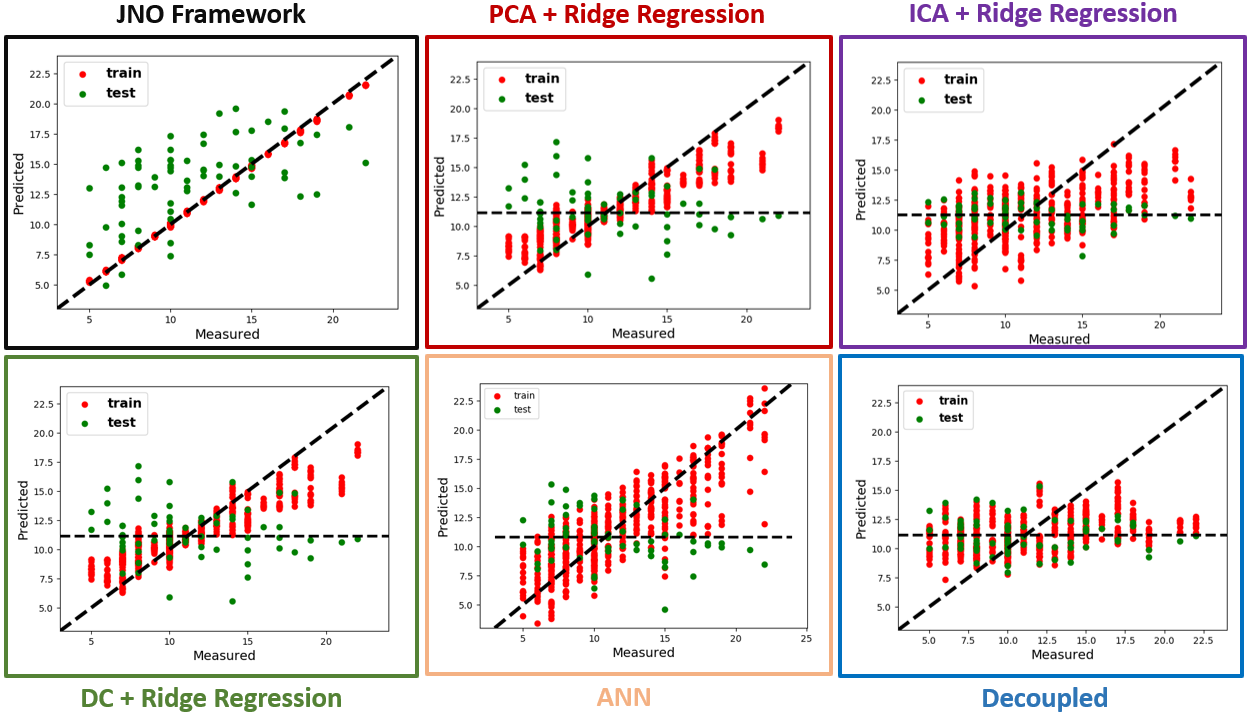}
   \caption{\textbf{NYU dataset:} Prediction performance for the ADOS score for \textbf{Black Box:} JNO Framework. \textbf{Red Box:} PCA and ridge regression \textbf{Purple Box:} ICA and ridge regression \textbf{Green Box:} Node degree centrality and ridge regression \textbf{Orange}: ANN on correlation features \textbf{Blue Box:} Decoupled matrix factorization and ridge regression}
   \label{fig6:Fig6 ABIDE ADOS}
\end{figure*}

\begin{figure*}    
    \centering
      \includegraphics[width=\dimexpr \textwidth-2\fboxsep-2\fboxrule\relax, scale=0.8]{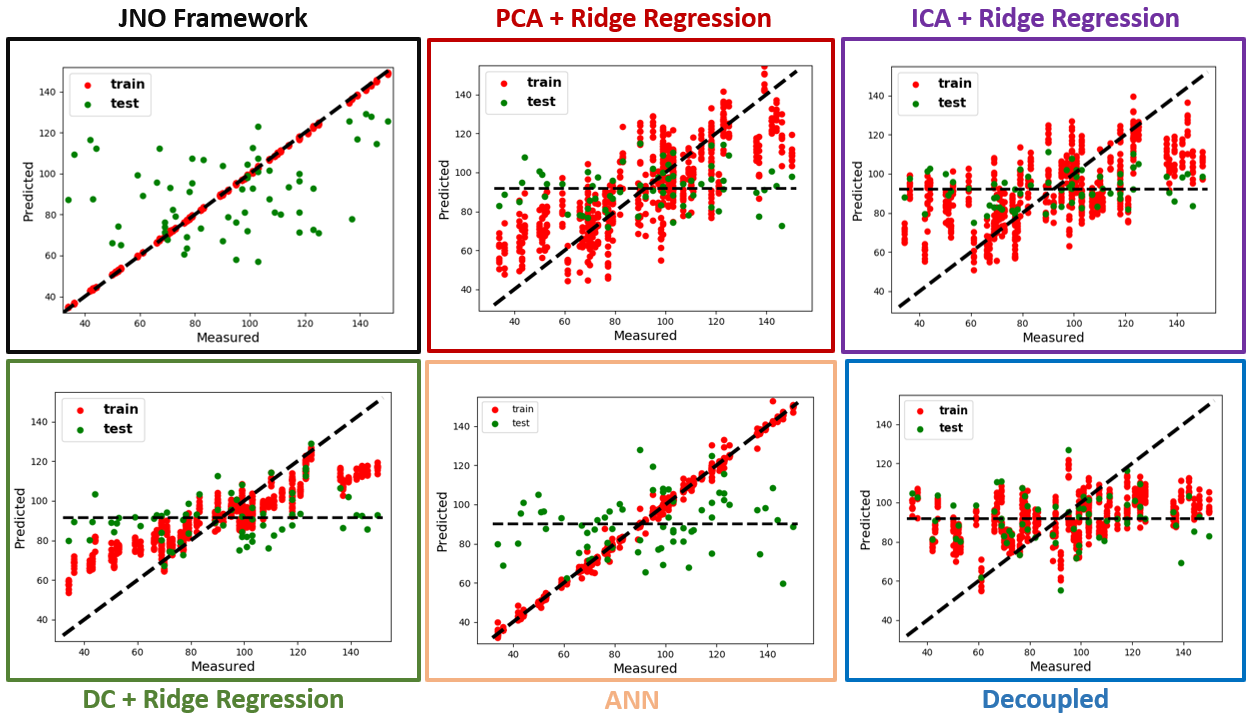}

  \caption{\textbf{NYU dataset:} Prediction performance for the SRS score for \textbf{Black Box:} JNO Framework. \textbf{Red Box:} PCA and ridge regression \textbf{Purple Box:} ICA and ridge regression \textbf{Green Box:} Node degree centrality and ridge regression \textbf{Orange}: ANN on correlation features \textbf{Blue Box:} Decoupled matrix factorization and ridge regression} 
  \label{fig7:Fig7 ABIDE-SRS}
\end{figure*}
\begin{table*}[t!]
\centering
{\renewcommand{\arraystretch}{1.2}
\begin{tabular}{|c |c | c| c| c| c | c |} 
\hline 
  \textbf{Score} &\textbf{Method} &\textbf{MAE Train } & \textbf{MAE Test} & \textbf{NMI Train} & \textbf{NMI Test} & \textbf{CDF-KS} \\  
\hline 
\hline
  \multirow{6}{4em}{ADOS} &PCA \& ridge & 2.18~\rpm~2.2 & 2.99~\rpm~{1.71} & 0.22 & 0.18 & \underline{0.06}\\
  & ICA \& ridge & 2.13~\rpm~1.1 & 3.01~\rpm~{1.90} & 0.31 & 0.23 & \textbf{0.027}\\
 & $D_{N}$ \& ridge & 1.22~\rpm~0.91 & 3.68~\rpm~{2.53} & 0.45& 0.39 & \textbf{0.015}\\
 & ANN  & 2.68~\rpm~2.21 & \textbf{2.28~\rpm~{1.30}} & \underline{0.91}& \textbf{0.58} & 0.088\\
 & Decoupled & 2.36~\rpm~2.33 & 2.63~\rpm~{1.90}& 0.15& 0.30 & 0.083\\
 & \textbf{JNO Framework} & \textbf{0.088~\rpm~0.13}& \underline{2.53~\rpm~{1.86}}& \textbf{0.99}& \underline{0.52} & {$-$}\\
[0.2ex]  
\hline
 \multirow{6}{4em}{SRS} &PCA \& ridge & 12.92~\rpm~10.48 & 19.09~\rpm~{12.48} & 0.64 & 0.39 & \textbf{0.032}\\
  & ICA \& ridge & 7.96~\rpm~6.35 & 20.8~\rpm~{17.3} & 0.83 & 0.63 & \textbf{0.041}\\
 & $D_{N}$ \& ridge & 5.77~\rpm~4.88 & 19.63~\rpm~{17.23} & 0.85& 0.59 & 0.089\\
 & ANN  & 4.77~\rpm~4.09 & 21.25~\rpm~{14.63} & 0.81 & 0.56 & 0.093\\
 & Decoupled & 12.06~\rpm~10.04 & 18.5~\rpm~{16.4}& 0.74& 0.37 & \textbf{0.014}\\
 & \textbf{JNO Framework} & \textbf{0.13~\rpm{0.07}}& \textbf{13.27~\rpm~{10.85}}& \textbf{0.99}& \textbf{0.78} & {$-$}\\ [0.2ex]
 \hline
 \multirow{6}{4em}{Praxis} &PCA \& ridge & 9.44~\rpm~6.83 & 12.83~\rpm~{8.84} & 0.64 & 0.37 & 0.17\\
  & ICA \& ridge & 4.79~\rpm~4.17 & 13.08~\rpm~{13.07} & 0.73 & 0.63 & \textbf{0.035}\\
 & $D_{N}$ \& ridge & 4.78~\rpm~3.24 & 13.93~\rpm~{8.14} & 0.68& 0.56 & \textbf{0.017}\\
 & ANN  & 9.34~\rpm~7.21 & 14.90~\rpm~{10.06} & 0.69& 0.39 & \textbf{0.01}\\
 & Decoupled & 10.17~\rpm~7.96 & 13.24~\rpm~{10.38}& 0.68& 0.29 & 0.10\\
 & \textbf{JNO Framework} & \textbf{0.11~\rpm~{0.065}}& \textbf{10.18\rpm~{6.58}}& \textbf{0.99}& \textbf{0.79} & $-$\\ [0.2ex]
 \hline
\end{tabular}
}
\caption{\textbf{KKI Dataset:} Performance evaluation using \textbf{Median Absolute Error (MAE)} and \textbf{Normalized Mutual Information (NMI)} fit, both for testing \& training. Lower MAE \& higher NMI score indicate better performance. We have highlighted the best performance in bold. The \textbf{CDF-KS} column indicates the Kolmogorov-Smirnoff statistic on the CDF comparison against our method. The instances highlighted in bold indicate that the performance is within the accepted $0.05$ threshold. Near misses have been underlined.}
\label{table:1}
\end{table*} 

\begin{table*}
\centering
{\renewcommand{\arraystretch}{1.2}
\begin{tabular}{|c |c | c| c| c| c| c|} 
\hline 
  \textbf{Score} &\textbf{Method} &\textbf{MAE Train} & \textbf{MAE Test} & \textbf{NMI Train} & \textbf{NMI Test} & \textbf{CDF-KS}\\  
\hline 
\hline
  \multirow{6}{4em}{ADOS} &PCA \& ridge & 1.68~\rpm~{1.53} & 3.46~\rpm~{2.21} & 0.30 & 0.28 & \textbf{0.0019}\\
 & ICA \& ridge & 2.75~\rpm~{1.88} & 3.41~\rpm~{2.34}& 0.15& 0.17 & \underline{0.068}\\
 & $D_{N}$ \& ridge & 1.18~\rpm~{1.19} & 3.17~\rpm~{3.05}& 0.50& 0.39 & \textbf{0.0025}\\
 & ANN  & 1.40~\rpm~1.39 & 3.36~\rpm~{2.89} & 0.31 & 0.28 & 0.081\\
 & Decoupled & 2.62~\rpm~2.54 & 3.32~\rpm~{2.27}& 0.21 & 0.12 & \underline{0.07}\\
 & \textbf{JNO Framework} & \textbf{0.10~\rpm~{0.088}}& \textbf{2.63~\rpm~{2.51}}& \textbf{0.99}& \textbf{0.54} & $-$\\
[0.2ex]  
\hline
 \multirow{6}{4em}{SRS} &PCA \& ridge & 10.64~\rpm~{12.60} & 18.22~\rpm~{12.43} & 0.87 & 0.54 & 0.24\\ [0.2ex]  
  & ICA \& ridge & 16.88~\rpm~{15.71} & 18.11~\rpm~{13.9}& 0.71& 0.49 & 0.19\\
 & $D_{N}$ \& ridge & 7.32~\rpm~{5.76} & 23.18~\rpm~{18.44} & 0.87& 0.66 & 0.081\\
 & ANN  & 1.50~\rpm~1.39 & 19.04~\rpm~{17.69} & 0.85 & 0.08 & \textbf{0.009}\\
 & Decoupled & 16.42~\rpm~15.66 & 22.43~\rpm~{18.79}& 0.80 & 0.42 & \underline{0.06}\\
 & \textbf{JNO Framework} & \textbf{0.46~\rpm~{0.36}}& \textbf{16.61~\rpm~{12.43}}& \textbf{0.96}& \textbf{0.72} & $-$\\ [0.2ex]
 \hline
\end{tabular}
}
\caption{\textbf{NYU Dataset:} Performance evaluation using \textbf{ Median Absolute Error (MAE)} and \textbf{Normalized Mutual Information (NMI)} fit, both for testing \& training. Lower MAE \& higher NMI score indicate better performance. We have highlighted the best performance in bold. The \textbf{CDF-KS} column indicates the Kolmogorov-Smirnoff statistic on the CDF comparison against our method. The instances highlighted in bold indicate that the performance is within the accepted $0.05$ threshold. Near misses have been underlined.}
\label{table:2}
\end{table*} 
Figs.~\ref{fig6:Fig6}$-$\ref{fig8:Fig8} compare the performance of our method against the baselines described in Section~\ref{baselines} for the prediction of ADOS, SRS and Praxis respectively for the KKI dataset. Similarly, Fig.~\ref{fig6:Fig6 ABIDE ADOS} and Fig.~\ref{fig7:Fig7 ABIDE-SRS} illustrate the performance comparison on the NYU dataset for ADOS and SRS respectively. We plot the score predicted by the algorithm on the $y$-axis against the measured ground truth score on the $x$-axis. The bold $x=y$ line indicates ideal performance. The red points correspond to training data, while the green points represent the held out testing data for all the folds in the cross validation. Our method is indicated at the top left corner in each plot. We observe that, although the training performance of the baselines is good (i.e. the red points follow the $x=y$ line), the JNO achieves the best training performance in all cases. Furthermore, we notice that all the two stage baseline testing performances track the mean value of the held out data (indicated by the black horizontal line). Our method clearly outperforms the baselines and is able to capture a trend in the data, beyond a mean value estimation in case of both datasets for all scores. This can be observed by the spread of the green points about the $x=y$ line in the case of the JNO method. Through our experiments, we noticed that the testing performance of the ANN is dependent on the choice of architecture. For example, the architecture chosen in Section~\ref{baselines} performs well on predicting ADOS for the KKI dataset, but performs poorly on all other comparisons. Our empirical evaluations could not identify a single architecture that performed well in all cases, like our JNO framework. The failure of the two stage decomposition in the bottom right comparison figures strengthens our hypothesis that a joint modeling of the neuroimaging and behavioral data is necessary in the context of generalization onto unseen data.
\par The lackluster generalization performance of the baselines is testament to the difficulty of the task at hand. The number of connections or features available to us are of the order of a $6670$ dimensional vector representation for $58$ or $63$ patients. Both the machine learning and graph theoretic techniques we selected for a comparison are well known in literature for being able to robustly provide compact characterizations for high dimensional datasets. However, we see that PCA and ICA are unable to estimate a reliable projection of the data that is particularly indicative of clinical severity. Similarly, the node degree measure heavily rely on being able to accurately identify informative network topologies from the observed correlation matrices. However, its aggregate nature captures general trends and is not successful in characterizing subtle patient level differences. The failure of the decoupled matrix factorization and ridge regression makes a strong case for including the regression term as a part of our JNO objective. The basis directions obtained in this case are not indicative of clinical severity, due to which the regression performance suffers. Despite sweeping parameters across several orders of magnitude, we observe that the baselines are only good at capturing group level information, as is indicated by the training fit. However, they fail to characterize patient level differences for an unseen subject and simply predict the mean of the given cohort.
\par On the other hand, the generalization power of the ANN is contingent on the model order choice. This is demonstrated by its inability to perform well on comparisons outside of ADOS for the KKI dataset. Said another way, we have to change the network architecture for different severity measures across datasets. This is a major computational disadvantage when compared with our method.
\par A key difference between the JNO framework and the baselines is that we utilize the structure of the correlation matrices to guide the predictive model. In essence, we optimize for the tradeoff between the neuroimaging and behavioral data representations jointly, instead of posing it as a two stage problem. The matrix decomposition we employ explicitly models the group information through the basis, and the patient differences through the coefficients. The limited number of basis elements we employ to decompose the data provides us with compact representations which explain the connectivity information well. The regularization terms and constraints ensure that the problem is well posed, while providing clinically meaningful and informative representations about the data. We also quantify the performance indicated in these figures in Tables~\ref{table:1} (KKI dataset) and Table~\ref{table:2} (NYU dataset) based on the validation metrics mentioned earlier.
\subsection{Subnetwork Identification}
\begin{figure*}[t]
   \centering
   \includegraphics[width=\dimexpr \textwidth-2\fboxsep-2\fboxrule\relax, scale=0.8]{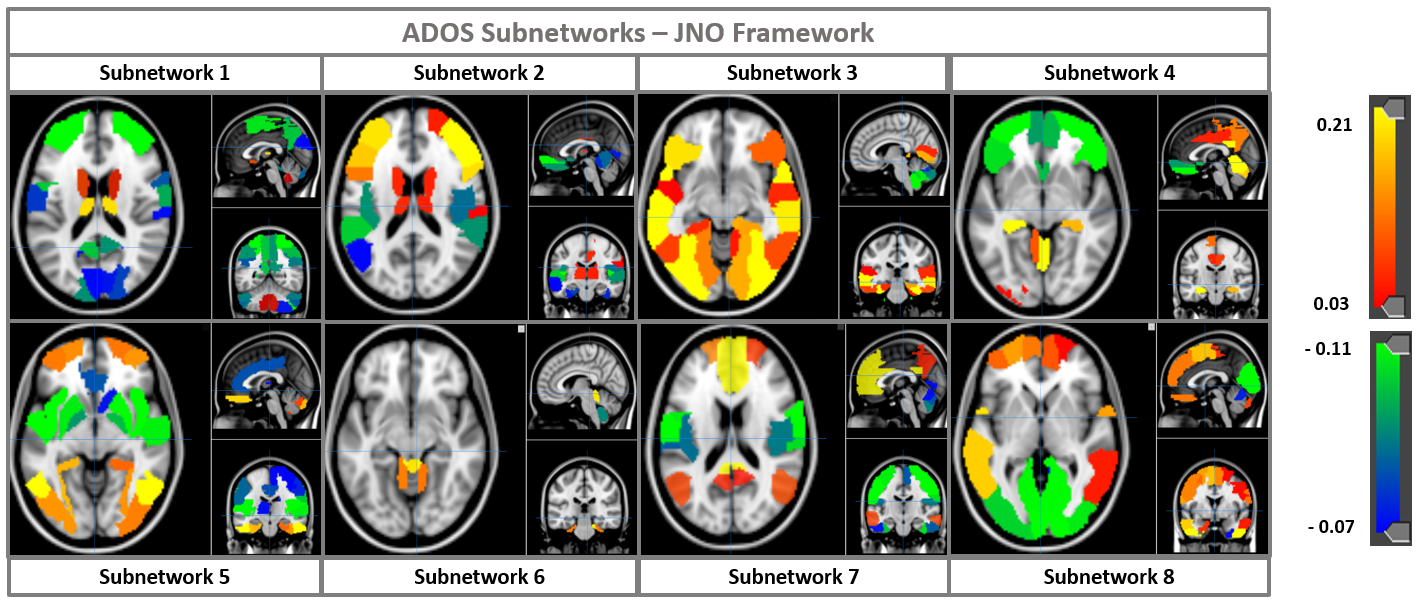}
   \caption{Subnetworks estimated to predict the ADOS score by the JNO. Regions having negative contributions are anti-correlated with areas having positive values} \label{fig11:Fig11}
\end{figure*}

\begin{figure*}
   \centering
   \includegraphics[width=\dimexpr \textwidth-2\fboxsep-2\fboxrule\relax, scale=0.8]{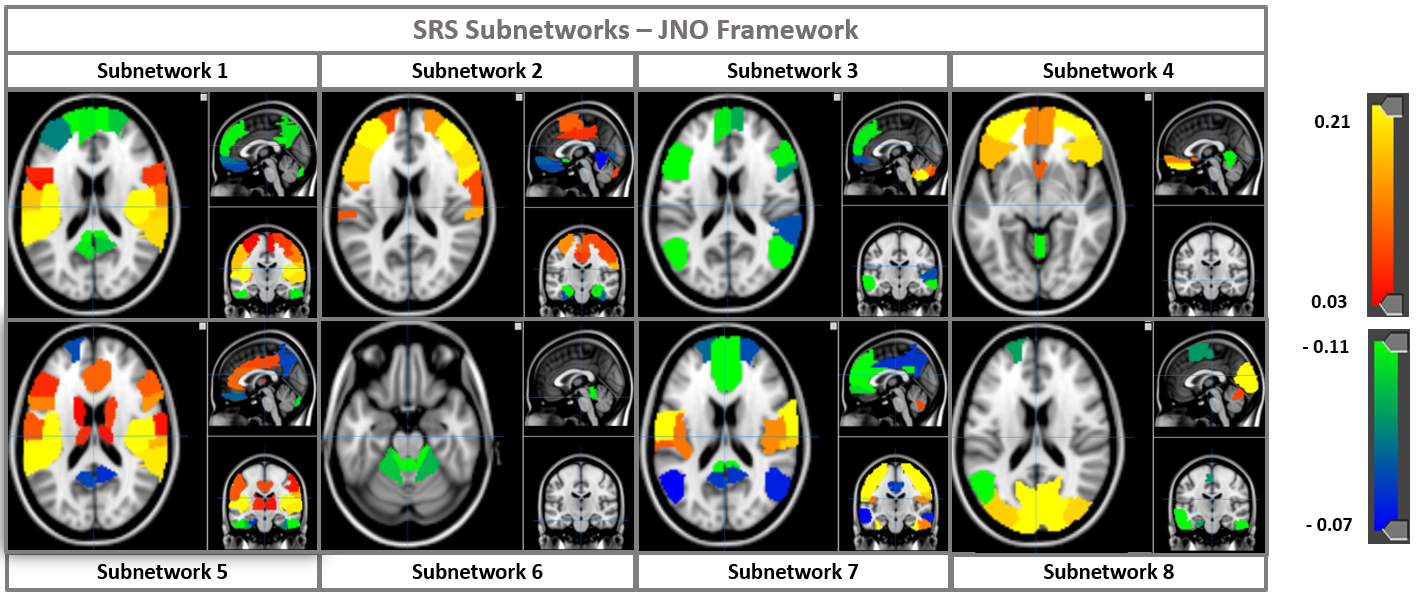}
   \caption{Subnetworks estimated to predict the SRS score. Regions having negative contributions are anti-correlated with areas having positive values} \label{fig12:Fig12}
\end{figure*}
\begin{figure*}
   \centering
   \includegraphics[width=\dimexpr \textwidth-2\fboxsep-2\fboxrule\relax, scale=0.8]{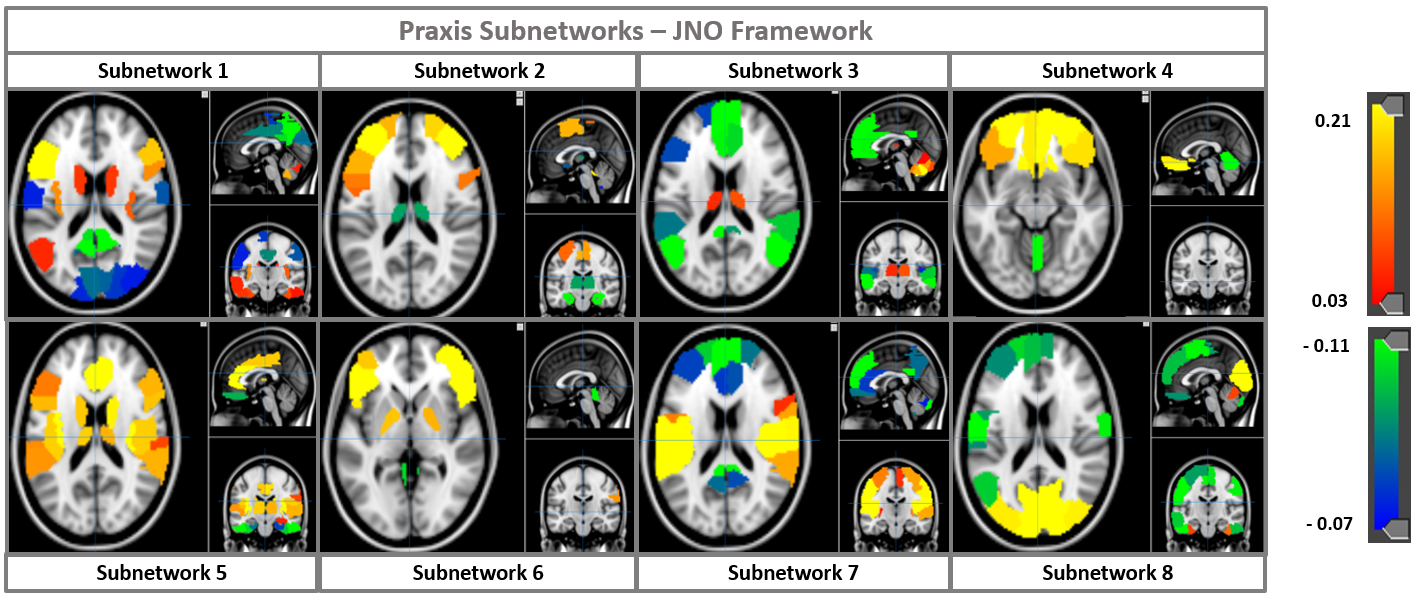}
   \caption{Subnetworks estimated to predict the Praxis score. Regions having negative contributions are anti-correlated with areas having positive values} \label{Praxissub}
\end{figure*}
\begin{figure}[t!]
 \centering
   \includegraphics[scale=0.35]{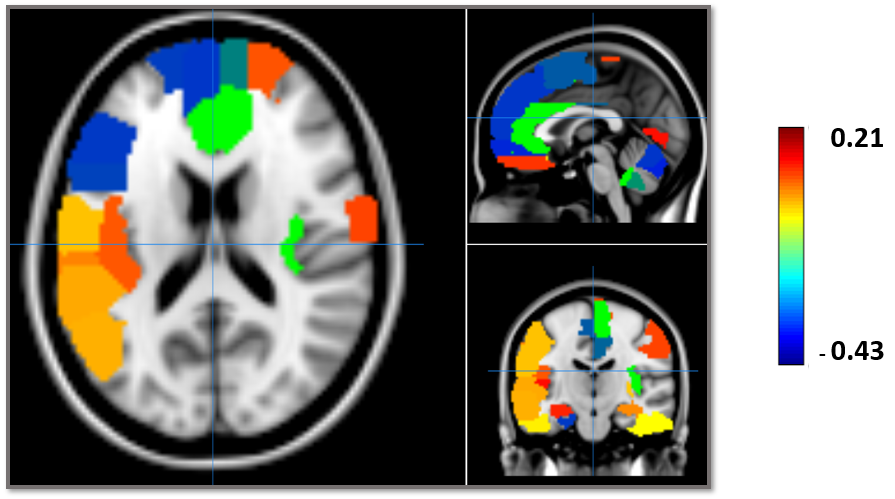}
   \caption{Representation learned from the prediction of ADOS by Node degree centrality + ridge regression. The colorbar indicates the weight of the ROI assigned by the ridge regression}  \label{DC_SubN}

\end{figure}
\begin{figure}[t!]
   \centering
   \includegraphics[scale=0.43]{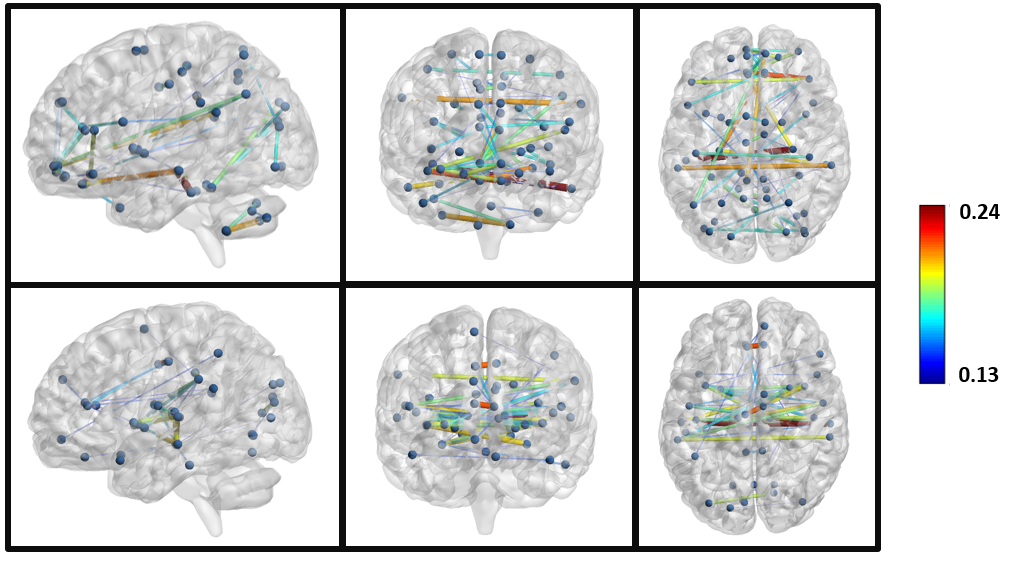}
   \caption{Top two subnetworks identified by the prediction of the ADOS score by PCA + ridge regression. The colorbar indicates the weight of the connection. } \label{PCA_SubN}
\end{figure}
\begin{figure}[b!]
   \centering
   \includegraphics[scale=0.4]{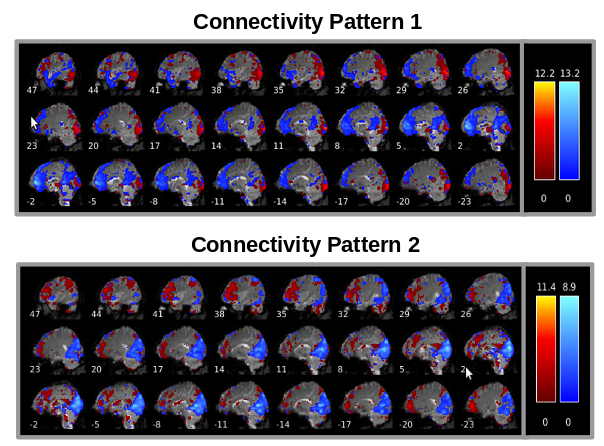}
   \caption{Connectivity patterns identified as important in the prediction of the ADOS score by ICA + ridge regression. Each plot displays $2$ spatial components contributing to the correlation feature. The colorbar indicates the weight of the connection.  } \label{ICA_SubN}
\end{figure}
\begin{figure}[b!]
   \centering
   \includegraphics[scale=0.43]{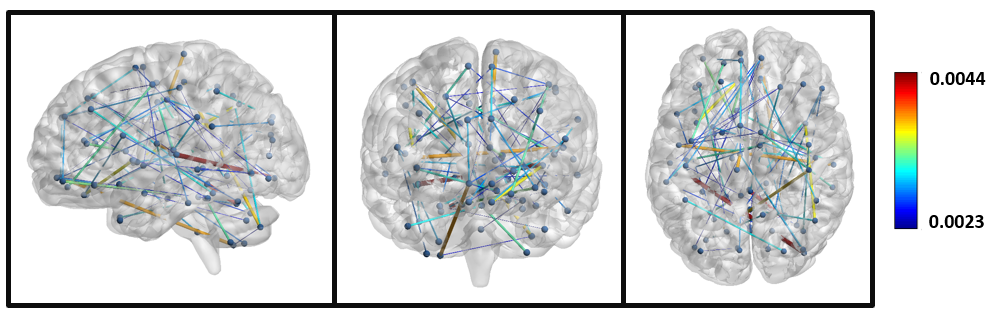}
   \caption{Connectivity patterns identified in the prediction of the ADOS score by the ANN. The colorbar indicates the weight of the connections. The narrow range of values are indicative that the ANN assigns equal weighting to most connections on an average } \label{ANN_SubN}
\end{figure}
\begin{figure*}[t]
   \centering
   \includegraphics[width=\dimexpr \textwidth-2\fboxsep-2\fboxrule\relax, scale=0.8]{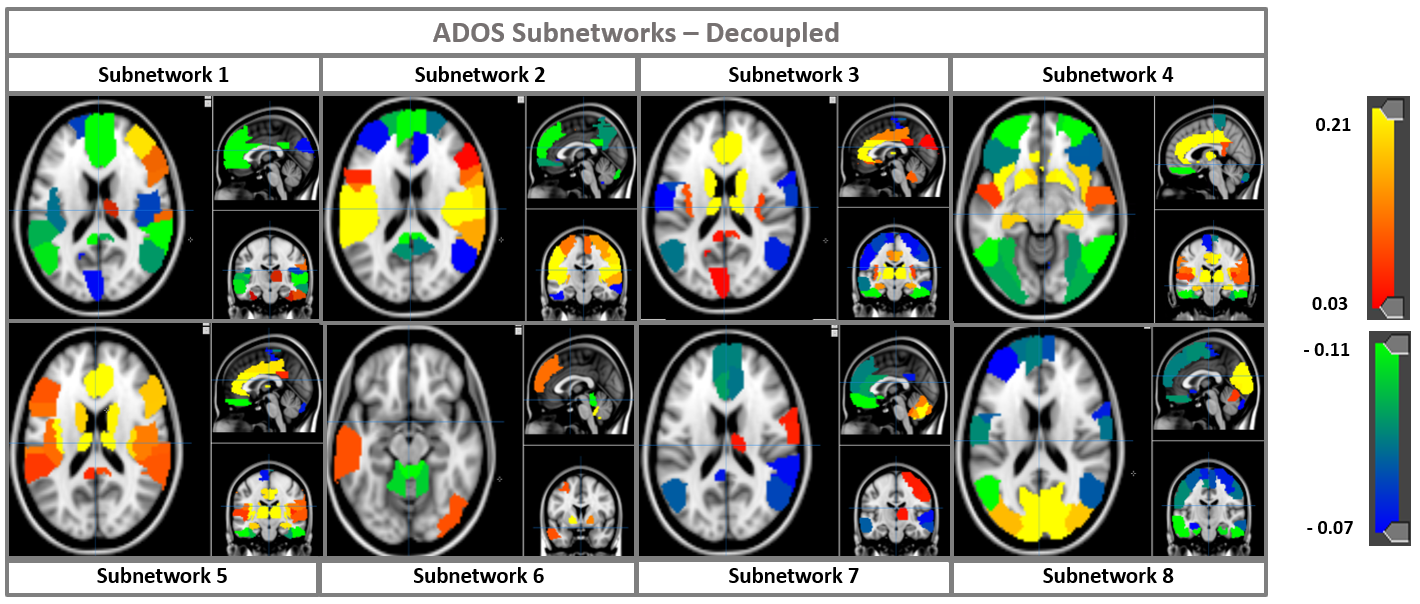}
   \caption{Subnetworks estimated to predict ADOS score by decoupling the matrix decomposition and ridge regression. Regions having negative contributions are anti-correlated with areas having positive values} \label{Decoupled}
\end{figure*}

Figs.~\ref{fig11:Fig11}$-$\ref{Praxissub} illustrate the subnetworks in $\mathbf{B}$, as trained on the ADOS, SRS and Praxis in the KKI dataset, respectively. Since each column of the basis corresponds to a set of co-activated subregions, we plot the normalized values stored in these columns onto the corresponding AAL ROIs. The colorbar indicates subnetwork contribution to the AAL regions. Regions colored as negative values are anticorrelated with regions storing positive ones. We rank the $8$ subnetworks obtained from SRS and Praxis according to their overlap with the subnetworks from ADOS. As seen from these figures, corresponding subnetworks show considerable overlap in regional co-activation patterns. The individual variations can arise from the fundamental differences in the behavioral traits that each score is trying to capture.
\par From a clinical standpoint, Subnetwork~$7$ includes competing i.e. anticorrelated contributions from regions of the default mode network (DMN) and somatomotor network (SMN). Abnormal connectivity within the DMN and SMN has been previously reported in ASD [\cite{lynch2013default,nebel2016intrinsic}]. Subnetwork~$5$ comprises of competing contributions from SMN regions. Additionally, it includes higher order visual processing areas in the occipital and temporal lobes, which is consistent with behavioral reports of reduced visual-motor integration in the ASD literature [\cite{nebel2016intrinsic}]. Subnetwork~$1$ has competing  from prefrontal and subcortical contributions, mainly the thalamus, amygdala and hippocampus. The thalamus is responsible for relaying sensory and motor signals to the cerebral cortex in the brain. The hippocampus is known to play a crucial role in the consolidation of long and short term memory, along with spatial memory to aid navigation. Altered memory functioning has been shown to manifest in children diagnosed with ASD [\cite{williams2006profile}]. Along with the amygdala, which is known to be associated with emotional responses, these areas may be crucial for social-emotional regulation in ASD.  Finally, Subnetwork~$2$ is comprised of competing contributions from the central executive control network and the insula, which is thought to be critical for switching between self-referential and goal-directed behavior [\cite{sridharan2008critical}]. 
\subsubsection{Robustness in Subnetwork Recovery}
Notice that we estimate a different basis matrix $\mathbf{B}$ for each cross validation fold. Therefore, one important property to verify is that these subnetworks are similar across different cohorts of the data.
\par We observed an average similarity of $0.79 \rpm 0.06$ for the ADOS networks, $0.86 \rpm 0.04$ for the SRS networks, and $0.76 \rpm 0.06$ for the Praxis networks across their cross validation runs. Additionally, upon a cross comparison between the ADOS and SRS networks, we obtained an average similarity of $0.82 \rpm 0.07$. Similarly, the overlap between ADOS and Praxis is $0.79 \rpm 0.04$, and between SRS and Praxis is $0.77 \rpm 0.06$. For a convenient visual inspection, we have arranged the networks in Fig.~\ref{fig12:Fig12} (SRS) and Fig.~\ref{Praxissub} (Praxis) in the order of their inner product similarity with the ADOS networks in Fig.~\ref{fig11:Fig11}. This finding strengthens the hypothesis that our model is successful at capturing the stable underlying mechanisms which explain the different sets of deficits of the disorder.

\subsubsection{Comparing Subnetwork Representations} In this section, we compare the subnetworks identified by  the JNO to the representations learned by the baseline methods. Recall that we have used a regularized linear regression as the Stage~$2$ predictor for the baselines. Therefore, we can probe the learned regression weights to characterize the baseline network representations.

\par Degree centrality looks at the relative importance of each brain region or `node' to the overall representation. To visualize the pattern identified by the degree centrality + ridge regression baseline, we display the regression weights on the brain surface plots in Fig.~\ref{DC_SubN}, normalized to unit norm. The colorbar indicates the strength of co-activation. Regions storing negative values are anticorrelated with regions storing positive weights. We again observe patterns from the DMN in the subnetwork plot. Note that the DMN was also a key connectivity pattern identified by the JNO. However, several other subnetworks identified by the JNO do not figure in this representation.

\par On the other hand, for the PCA + ridge regression baseline, the regression weights inform us of the relative importance of the principal components in prediction. Since the features fed into PCA are the $M = (P\times(P-1))/2$ correlation values, we are left with a $6670$ dimensional edge connectivity representation for the AAL per component. We first examine the absolute value of the regression weights learned, and then display the connectivity in the top $2$ basis components in Fig.~\ref{PCA_SubN}. We render this connectivity measure using the BrainNet Viewer [\cite{xia2013brainnet}] software. For clarity, we have chosen to display the top $5$ percent of the connections obtained. The solid edges signify retained connections, while the blue spheres correspond to nodes of the AAL regions. The colorbar to the right indicates the strength of the connections. We notice that the components consist of several crossing connections spread across different regions of the brain. As compared to our model, which pinpoints key subnetworks already known to be associated with ASD, the representation obtained is not immediately interpretable.

\par In the ICA + ridge regression baseline, the input to the regression model are the correlation values between the components identified by ICA. After the model is fit, we sort the input correlations based on the learned regression weights. This helps us identify the features important for prediction. In Fig.~\ref{ICA_SubN} , we display the spatial maps of the top $2$ connections identified by the algorithm. We again, observe patterns from the DMN and visual areas. However, it fails to capture several other subnetwork patterns that the JNO identifies as important for ASD.

\par For the ANN, we use the weight matrix learned at the input layer to inform us of the subnetwork connectivity. Recall that this matrix is of dimension $M \times D$, where $M = (P \times (P-1))/2 = 6670$ for the AAL atlas. For our application, $D$ is of width $8000$. We first take the absolute values of these weights, and then normalize the columns of this matrix to unit norm. We then average across the rows to obtain a single $6670$ dimensional edge-edge connectivity vector. Again, we use the BrainNet connectivity plots to display this information in Fig.~\ref{ANN_SubN}. We have chosen to display the top $1$ percent of the connections obtained. The solid edges signify retained connections, while the blue spheres correspond to nodes of the AAL regions. The colorbar to the right indicates the strength of the connections. We observe several overlapping connectivity patterns spread across the entire brain despite applying a stringent threshold. Additionally, the narrow range of values indicates that the ANN assigns nearly equal weight to all connections on an average. Similar to the PCA baseline, this representation is unable to capture interpretable connectivity patterns which explain behavior.

\par Finally, we examine the representation learned by performing the matrix decomposition and prediction separately, i.e. the decoupled case. Note that the learned basis matrix $\mathbf{B}$ follows the same interpretation as that of the JNO. We display the corresponding co-activation patterns in Fig.~\ref{Decoupled}. Again, the colorbar indicates the strength of activation of the AAL ROIs. Negative regions are anticorrelated with the positive regions. For convenience, we have ordered the $8$ subnetworks according to their similarity with the ADOS subnetworks identified in Fig.~\ref{fig11:Fig11}. Since we use the same matrix decompotion as the JNO, we observe several similarities in the learned representations. We also notice subtle differences in the patterns on account of the coupling with the predictive term in the JNO. We conjecture that these learned differences are what gives the JNO the leverage to generalize to unseen data.
\begin{figure*}[h!]
   \centering
   \includegraphics[width=\textwidth- 2pt]{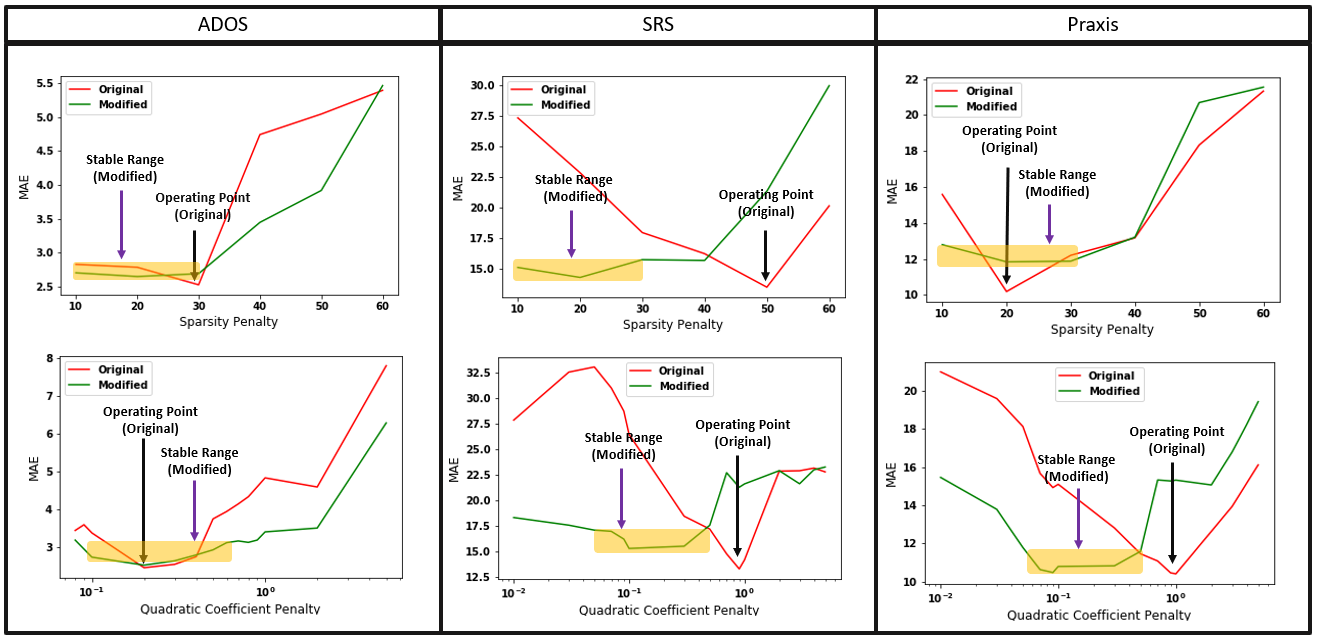}
   \caption{Comparing the sensitivity of the JNO framework with the modified objective in Eq.~(\ref{eqn23:Eqn23}). Prediction performance with varying \textbf{Top} $\lambda_1$ for  \textbf{(L-R):} ADOS, SRS and Praxis \textbf{Bottom} $\lambda_2$ for \textbf{(L-R):} ADOS, SRS and Praxis}
   \label{MitPar}
\end{figure*}

\subsection{Evaluating Model Generalizability} 
\label{MitiPara}

We have shown both predictive power and interpretabilitiy of our model thus far. Furthermore, characterizing model generalizability is important for future application of our framework. Here, we first examine the sensitivity of our prediction results with respect to the model hyperparameters. Then, we show robustness of our formulation to two common obstacles in generalizabilitiy for rs-fMRI analysis, namely choice of parcellation scheme and test-retest reliability. Lastly, we show mitigation strategies to handle hyperparameter sensitivity that make our framework more robust. 

\subsubsection{Hyperparameter Sensitivity}
As initially described in Section~\ref{Results}, our JNO framework is insensitive to the regression tradeoff $\gamma$ and ridge penalty $\lambda_{3}$. We also have a natural way to set the number of subnetworks $K$. However, we observe that our JNO framework is fairly sensitive to the sparsity on $\mathbf{B}$ and the ridge penalty on the coefficients $\mathbf{c}_n$, i.e. $\lambda_{1}$ and $\lambda_{2}$ respectively. Fig.~\ref{MitPar} represents the MAE recovery performance of the algorithm for varying settings of $\lambda_1$ and $\lambda_2$, holding the remaining parameter settings constant when evaluated on the KKI dataset. The red plots in each case indicate the performance of the JNO framework. The green plots allude to our mitigation strategy which we will present later in Subsection~\ref{Mitigation}. The $\mathbf{x}$-axis denotes the parameter value, while the $\mathbf{y}$-axis quantifies the MAE from cross validation. Observe that the best $\lambda_1$ and $\lambda_2$ settings for the individual scores are different, i.e ADOS-$\{\lambda_1=30, \lambda_2=0.2\}$, SRS-$\{\lambda_1=50, \lambda_2=0.9\}$, and Praxis-$\{\lambda_1=20, \lambda_2=0.6\}$. Additionally, the kinks in the plots (shown by the black arrow) also indicate that small changes in the sparsity and coefficient regularization lead to a dramatic change in performance, i.e. the operating points for these two parameters are narrow. We suspect that the hyperparameter differences can be partially attributed to the different dynamic range of each clinical score. Specifically, these differences impact the tradeoff between the representation learning and prediction terms in the JNO optimization. This in turn affects the generalization performance at a particular hyperparameter setting. These observations further illustrate the difficulty of the problem we are trying to address. Keeping these subtleties in mind, the next subsection focuses exclusively on studying the generalization performance of the JNO and the baselines at the optimal hyperparameter settings we obtained in Section~\ref{Results}.

\subsubsection{Examining Test-Retest Performance:}
Based on the observations above, we design two experiments to characterize the generalizability of the JNO with regards to its free parameters, i.e. $\lambda_{1}$ and $\lambda_{2}$, in a test-retest setting. The first of these experiments investigates the impact of varying the rs-fMRI parcellation scheme, which not only changes the dimensionality of the input correlation matrices, but also changes the distribution of correlation measures. The second experiment is a cross site comparison between KKI and NYU where the same parcellation (i.e. AAL) is maintained. However, site differences like acquisition protocol and pre-processing, impact the distribution of the input correlations. Overall, these two experiments validate generalizabilitiy and robustness of our model, as we show good performance under sub-optimal parameter settings.
\begin{table*}[b]
\centering

\begin{tabular}{|c |c | c| c| c| c | c |} 
\hline 
  \textbf{Score} &\textbf{Method} &\textbf{MAE Train } & \textbf{MAE Test} & \textbf{NMI Train} & \textbf{NMI Test} & \textbf{CDF-KS} \\  
\hline 
\hline
  \multirow{6}{4em}{ADOS} & PCA \& ridge & 1.19~\rpm~1.54 & 3.11~\rpm~{2.99} & 0.30 & 0.22 & \underline{0.061}\\
  & ICA \& ridge & 1.48~\rpm~1.11 & 3.13~\rpm~{3.01}& 0.58 & 0.43 & 0.08\\
  & $D_{N}$ \& ridge & 2.26~\rpm~1.88 & \textbf{2.41~\rpm~{2.69}} & 0.53 & 0.33 & \underline{0.059} \\
  & Decoupled & 1.44~\rpm~1.33 & 3.15~\rpm~{2.96} & 0.41 & 0.32 & 0.23\\
  & \textbf{JNO Framework} & \textbf{0.10~\rpm~0.096}& \underline{2.72~\rpm~{2.34}}& \textbf{0.99}& \textbf{0.50} & {$-$}\\
[0.2ex]  
\hline
 \multirow{6}{4em}{SRS} & PCA \& ridge & 6.73 ~\rpm~{5.94} & 19.62~\rpm~{14.30} & 0.84 & 0.46 & 0.19\\ [0.2ex]  
 & ICA \& ridge & 8.44~\rpm~{7.19} & 17.43~\rpm~{11.17} & 0.80 &  0.55 & \textbf{0.02}\\
 & $D_{N}$ \& ridge & 7.11~\rpm~{6.24} & 24.43~\rpm~{19.18} & 0.85 & 0.62 & 0.083\\
 & Decoupled & 8.19~\rpm~{7.95} & 19.16~\rpm~{14.13}& 0.84 & 0.46 & 0.08 \\
 & \textbf{JNO Framework} & \textbf{2.11~\rpm~{1.19}}& \textbf{16.03~\rpm~{14.58}}& \textbf{0.91}& \textbf{0.72} & {$-$}\\ [0.2ex]
 \hline
 \multirow{6}{4em}{Praxis} & PCA \&ridge & 6.91~\rpm~{6.04} & 13.67~\rpm~{8.36} & 0.79 & 0.51 & \underline{0.063}\\ [0.2ex]  
 & ICA \& ridge & 7.96~\rpm~{5.99} & 13.12~\rpm~{9.36} & 0.80 & 0.51 & \textbf{0.012}\\  
 & $D_{N}$ \& ridge & 6.41~\rpm~{5.99} & 13.21~\rpm~{7.96}& 0.81 & 0.49 & 0.09 \\
 & Decoupled & 11.56~\rpm~{11.78} & 17.96~\rpm~{16.11}& 0.69 & 0.48 & \underline{0.068}\\
 & \textbf{JNO Framework} & \textbf{0.36~\rpm~{0.27}}& \textbf{11.93~\rpm~{9.05}}& \textbf{0.95}& \textbf{0.74} & $-$\\ [0.2ex]
 \hline
\end{tabular}

\caption{\small{\textbf{KKI Dataset on Brainetome parcellation:} Performance evaluation using \textbf{ Median Absolute Error (MAE)} and \textbf{Normalized Mutual Information (NMI)} fit, both for testing \& training. Lower MAE \& higher NMI score indicate better performance. We have highlighted the best performance in bold. The \textbf{CDF-KS} column indicates the Kolmogorov-Smirnoff statistic on the CDF comparison against our method. The instances highlighted in bold indicate that the performance is within the accepted $0.05$ threshold. Near misses have been underlined. The parameter settings used were the same as those for the KKI dataset from the main manuscript}}
\label{Mit_BT}

\end{table*}

\begin{figure}[b!]
   \centering
   \includegraphics[scale=0.34]{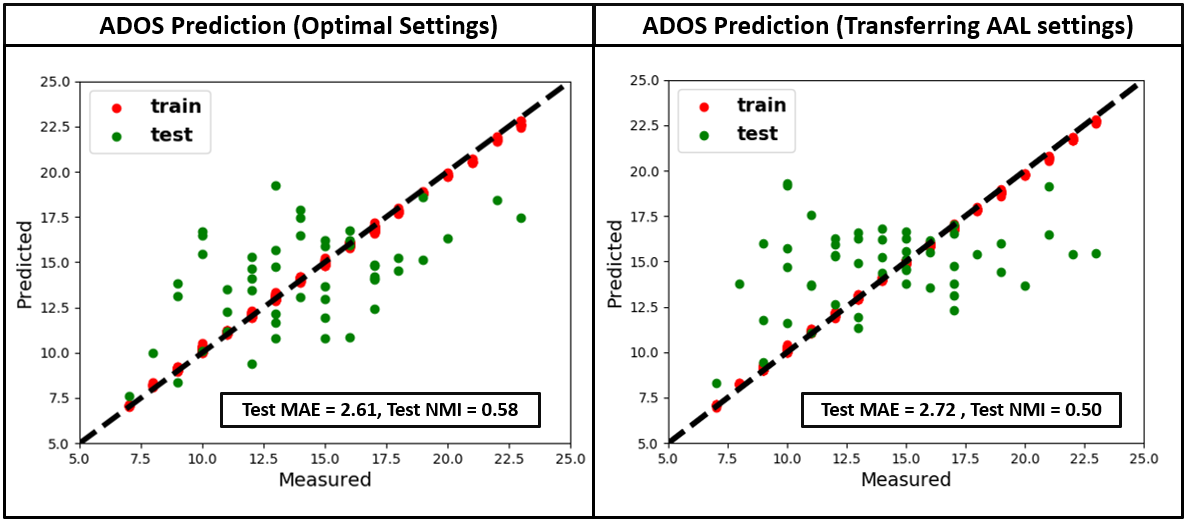}
   \caption{\textbf{Brainetome Parcellation:} A performance comparison for ADOS prediction by the JNO using the \textbf{(L)} Optimal Settings for Brainetome \textbf{(R)} Transferring AAL settings }\label{ADOS_Mit}
\end{figure}
\begin{figure}[b!]
   \centering
   \includegraphics[scale=0.335]{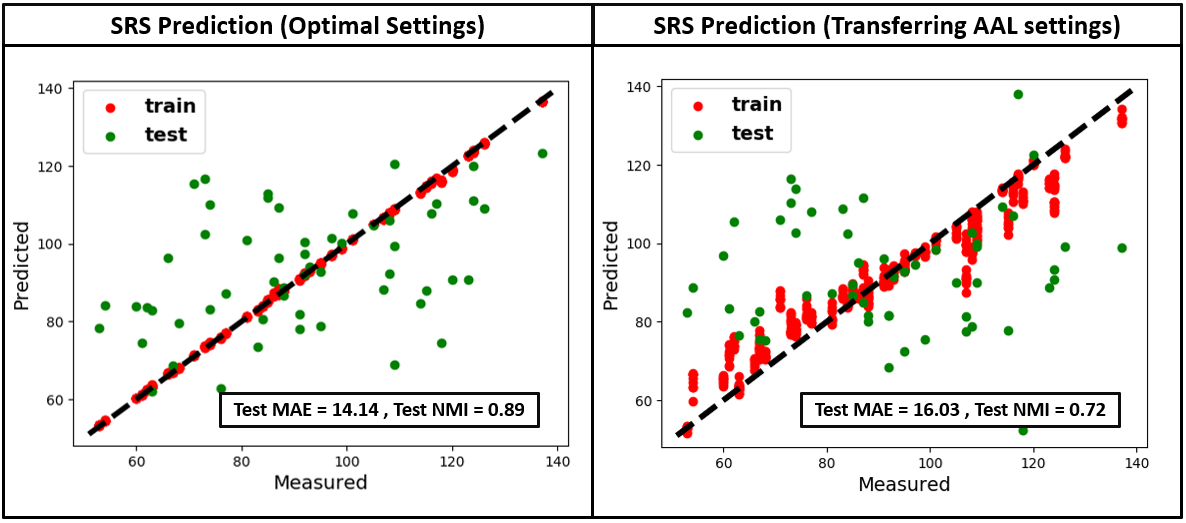}
   \caption{\textbf{Brainetome Parcellation:} A performance comparison for SRS prediction by the JNO using the \textbf{(L)} Optimal Settings for Brainetome \textbf{(R)} Transferring AAL settings}
   \label{SRS_Mit}
\end{figure}
\begin{figure}[b!]
   \centering
   \includegraphics[scale=0.34]{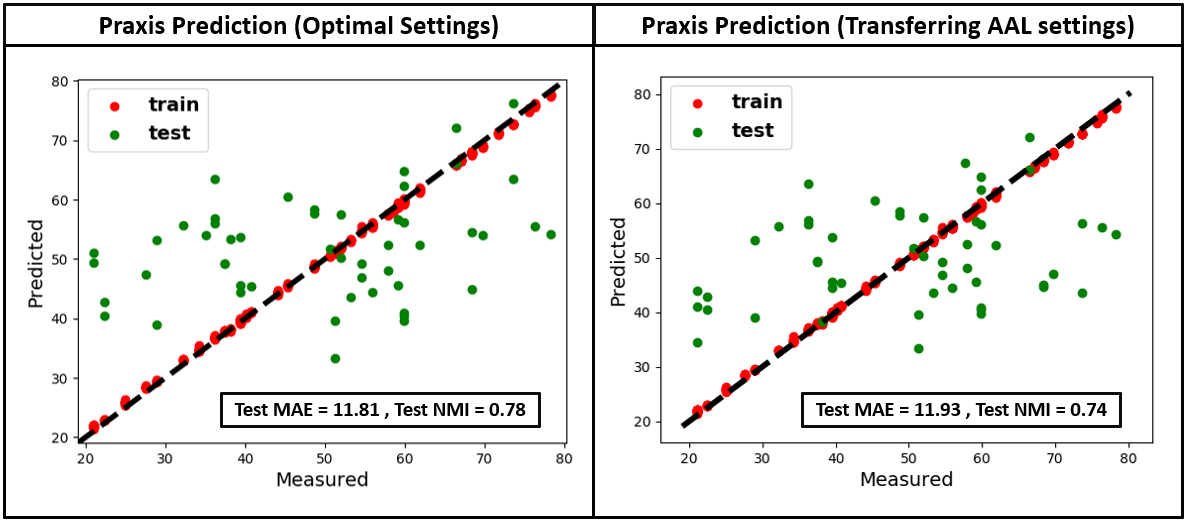}
   \caption{\textbf{Brainetome Parcellation:} A performance comparison for Praxis prediction by the JNO using the \textbf{(L)} Optimal Settings for Brainetome \textbf{(R)} Transferring AAL settings }\label{Praxis_Mit}
\end{figure}

\par For the first experiment, we compute correlation matrices for the KKI rs-fMRI dataset using the Brainetome-246 atlas. Note that the Brainetome [\cite{fan2016human}] provides a much finer ROI resolution as compared to the AAL-116 atlas used previously. We then predict the clinical scores from the new correlation matrices using the AAL hyperparameter settings. This is repeated for all the baselines along with the JNO. The plots in Figs.~\ref{ADOS_Mit}-\ref{Praxis_Mit} illustrate the prediction of ADOS, SRS and Praxis respectively by the JNO. In each figure, the left plots denote the Brainetome performance using the hyperparameter selection outlined in Section~\ref{Results}, while the right plots indicate the performance after transferring the hyperparameter settings from the AAL experiments. Notice that the performance is slightly worse than the case where they were explicitly set by cross validation (left plots). This is expected since we have not explicitly optimized for the new data characteristics. Table~\ref{Mit_BT} provides the corresponding quantitative metrics. We emphasize that the performance trends are very similar to those in Section~\ref{Results}, namely, we outperform nearly all the baselines, while retaining the interpretability of the subnetwork decomposition. Since the ROI resolution differs here, we can treat this comparison as a surrogate for out of sample generalization. Additionally, this experiment suggests that the JNO is agnostic to the parcellation scheme used. We have included additional baseline comparisons in the supplementary results document. We believe that the joint optimization helps us balance the tradeoff between the representation learning and prediction terms. Said another way, the matrix factorization term regularizes the problem and helps provide stability while handling the changes in the input data distribution. As a result, we are able to transfer the learned hyperparameter settings, yet outperform the baselines at severity prediction.

\begin{table*}[b!]
\centering
\renewcommand{\arraystretch}{1.2}
\begin{tabular}{|c |c | c| c| c| c| c|} 
\hline 
  \textbf{Score} &\textbf{Method} &\textbf{MAE Train} & \textbf{MAE Test} & \textbf{NMI Train} & \textbf{NMI Test} & \textbf{CDF-KS}\\  
\hline 
\hline
  \multirow{6}{4em}{ADOS} &PCA \& ridge & 1.51~\rpm~{1.49} & 3.52~\rpm~{3.21} & 0.41 & 0.23 & \textbf{0.039}\\
 & ICA \& ridge & 2.15~\rpm~{2.16} & 3.52~\rpm~{2.91}& 0.17 & 0.11 & \underline{0.07}\\
 & $D_{N}$ \& ridge & 2.16~\rpm~{2.39} & 3.91~\rpm~{3.05}& 0.39 & 0.31 & \underline{0.052}\\
 & ANN  & 1.40~\rpm~1.39 & 3.36~\rpm~{2.89} & 0.31 & 0.28 & \textbf{0.001}\\
 & Decoupled & 2.61~\rpm~2.13 & 3.51~\rpm~{3.17} & 0.29 & 0.09 & 0.21\\
 & \textbf{JNO Framework} & \textbf{0.08~\rpm~{0.06}}& \textbf{2.90~\rpm~{2.18}}& \textbf{0.99}& \textbf{0.41} & $-$\\
[0.2ex]  
\hline
 \multirow{6}{4em}{SRS} &PCA \& ridge & 11.19~\rpm~{14.16} & \textbf{16.25~\rpm~{14.11}} & 0.89 & 0.52 & 0.16\\ [0.2ex]  
  & ICA \& ridge & 16.98~\rpm~{16.62} & 18.90~\rpm~{15.14}& 0.73 & 0.43 & \textbf{0.041} \\
 & $D_{N}$ \& ridge & 8.91~\rpm~{6.51} & 23.52~\rpm~{16.10} & 0.88 & 0.51 & \underline{0.054}\\
 & ANN  & 1.50~\rpm~1.39 & 19.04~\rpm~{17.69} & 0.85 & 0.08 & 0.072\\
 & Decoupled & 15.11~\rpm~14.36 & 24.19~\rpm~{19.17}& 0.75 & 0.39 & \underline{0.059}\\
 & \textbf{JNO Framework} & \textbf{0.59~\rpm~{0.47}}& \underline{17.91~\rpm~{14.15}}& \textbf{0.95}& \textbf{0.65} & $-$\\ [0.2ex]
 \hline
\end{tabular}

\caption{\small{\textbf{NYU Dataset:} Performance evaluation using \textbf{ Median Absolute Error (MAE)} and \textbf{Normalized Mutual Information (NMI)} fit, both for testing \& training. Lower MAE \& higher NMI score indicate better performance. We have highlighted the best performance in bold. The \textbf{CDF-KS} column indicates the Kolmogorov-Smirnoff statistic on the CDF comparison against our method. The instances highlighted in bold indicate that the performance is within the accepted $0.05$ threshold. Near misses have been underlined. The parameter settings used were the same as those from the KKI dataset in the main manuscript}}
\label{Mit_NYU}
\end{table*}
\par
\begin{figure}[t!]
   \centering
   \includegraphics[scale=0.34]{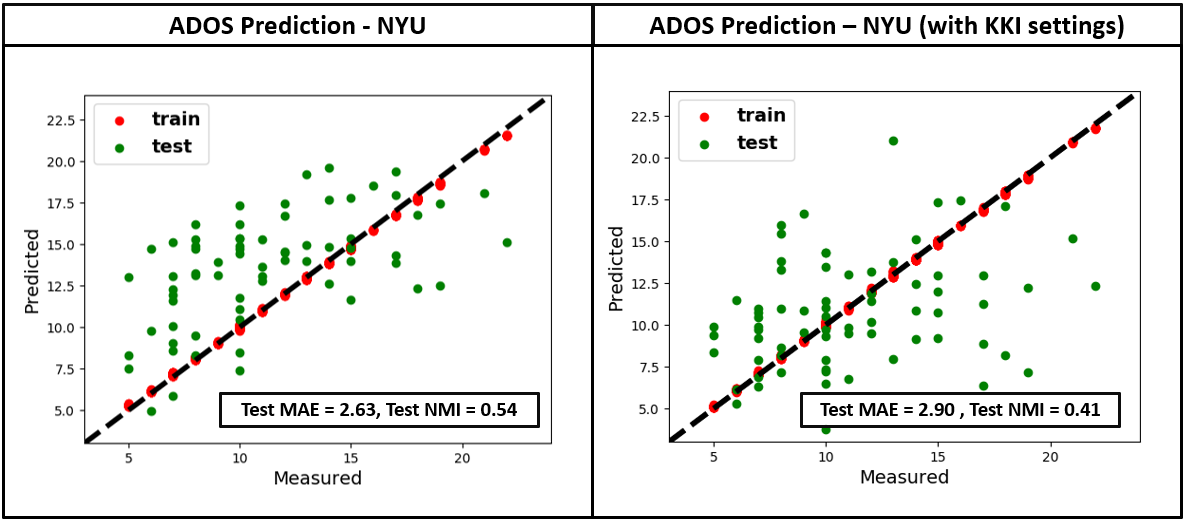}
   \caption{A performance comparison for ADOS prediction on the NYU Dataset by the JNO using the  settings learned from cross validation on the \textbf{(L)} NYU \textbf{(R)} KKI dataset.} \label{ADOS_NYU_Mit}
\end{figure}

\begin{figure}[t!]
   \centering
   \includegraphics[scale=0.34]{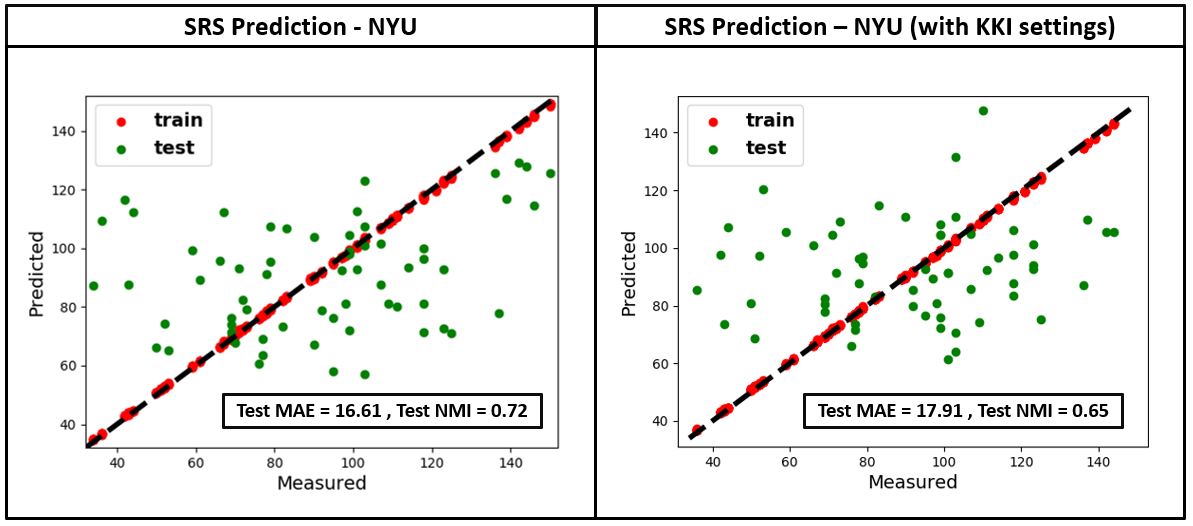}
   \caption{A performance comparison for SRS prediction on the NYU Dataset by the JNO using the  settings learned from cross validation on the \textbf{(L)} NYU \textbf{(R)} KKI dataset.}\label{SRS_NYU_Mit}
\end{figure}

 In our second experiment, we again use the hyperparameter settings learned from the KKI dataset, but this time, predict clinical scores from the NYU rs-fMRI correlation matrices. The parcellation scheme i.e. AAL, is maintained for both datasets. Figs.~\ref{ADOS_NYU_Mit} and \ref{SRS_NYU_Mit} compare these predictions (right sub-plots) with those from Section~\ref{Results} (left sub-plots) for ADOS and SRS respectively. Table~\ref{Mit_NYU} delineates the quantitative comparisons against the baselines. Again, we observe that the JNO still outperforms the baselines, though there is a slight reduction in quantitative performance. This reduction is expected, especially given the differences in scanning protocols across sites. As in the previous experiment, we are able to use the transferred parameter settings, and still optimize for a representation which is predictive of clinical severity. The cross dataset comparison is a good indicator of the robustness of the JNO to its hyperparameters. The observations from both experiments suggest that the JNO has the potential to capture robust and interpretable phenomenon related to the neurological disorder of interest. 
\par  Lastly, both the test-retest experiments study the overall generalizability of the method by examining a form of out of sample prediction, either by varying the parcellation scheme, or via a cross dataset comparison. This is very similar to the principles of nested cross validation, which aims to reduce bias in parameter selection.


\subsubsection{Mitigating Parameter Sensitivity}
\label{Mitigation}
Finally, we propose two main modifications to tackle the observed hyperparameter sensitivity in $\lambda_{1}$ and $\lambda_{2}$. Given that the dynamic ranges of the scores are quite different and potentially impact generalization, our first mitigation strategy is to rescale the measures to a fixed interval. Since ADOS is the most widely accepted observational measure of clinical autism severity, we have scaled and offset the remaining scores to have a range of $0$--$30$ (similar to ADOS).  To mitigate the narrow `operating point', we include an extra template average correlation term in Eq.~(\ref{eqn5:Eqn5}). We now model the residual outer-product terms as deviations around a mean template correlation matrix $\mathbf{B}_{avg}$. The rationale behind this additional term is that it encourage sparsity in the basis matrix along with the explicit $\ell_{1}$ penalty. The modified objective is as follows:
\begin{multline}
\mathcal{J}(\mathbf{B},\mathbf{B}_{avg},\mathbf{C},\mathbf{w}) = \sum_{n}{{ {\vert\vert{\mathbf{\Gamma}_{n} - \mathbf{B}_{avg} -\mathbf{B} \mathbf{diag}(\mathbf{c}_{n})}\mathbf{B}^{T} }\vert\vert}_{F}^{2} } \\ + \gamma{{\vert\vert{\mathbf{y} - \mathbf{C}^{T}\mathbf{w}}\vert\vert}}^{2}_{2} + \lambda_{1}{\vert\vert{\mathbf{B}}\vert\vert}_{1} \\ + \lambda_{2}{\vert\vert{\mathbf{C}}\vert\vert}^{2}_{F} + \lambda_{3}{\vert\vert{\mathbf{w}}\vert\vert}^{2}_{2} \ \ s.t. \ \ \mathbf{c}_{nk} \geq 0 ,
\label{eqn23:Eqn23}
\end{multline}

Notice that $\mathbf{B}_{avg}$ has a closed form update, which does not add much computational overhead. The updates for the remaining variables follow the same procedure as described in Section.~\ref{Optim}, except that the term, $\{\mathbf{\Gamma}_{n}\}$ is replaced with $\{\mathbf{\Gamma}_{n}-\mathbf{B}_{avg}\}$ in every update.
\par
\begin{figure}[t!]
   \centering
   \includegraphics[scale=0.34]{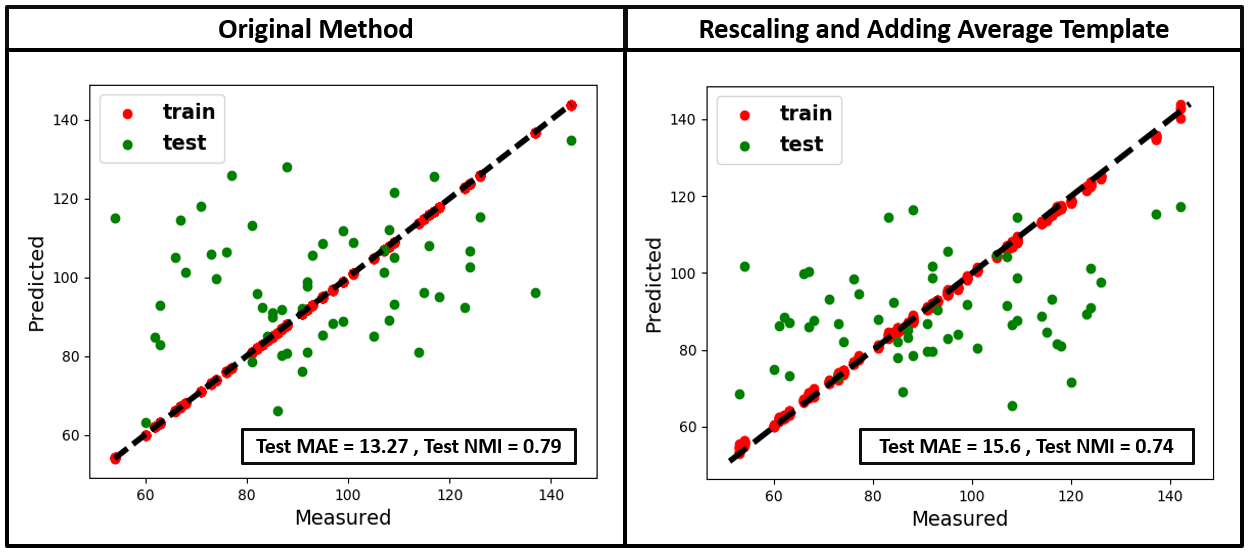}
   \caption{A performance comparison for SRS prediction after modifying the objective according to Eq.~(\ref{eqn22:Eqn22}). \textbf{(L)} Original Method \textbf{(R)} After re-scaling and average template addition}\label{fig9:Fig9}
\end{figure}

\begin{figure}[t!]
   \centering
   \includegraphics[scale=0.34]{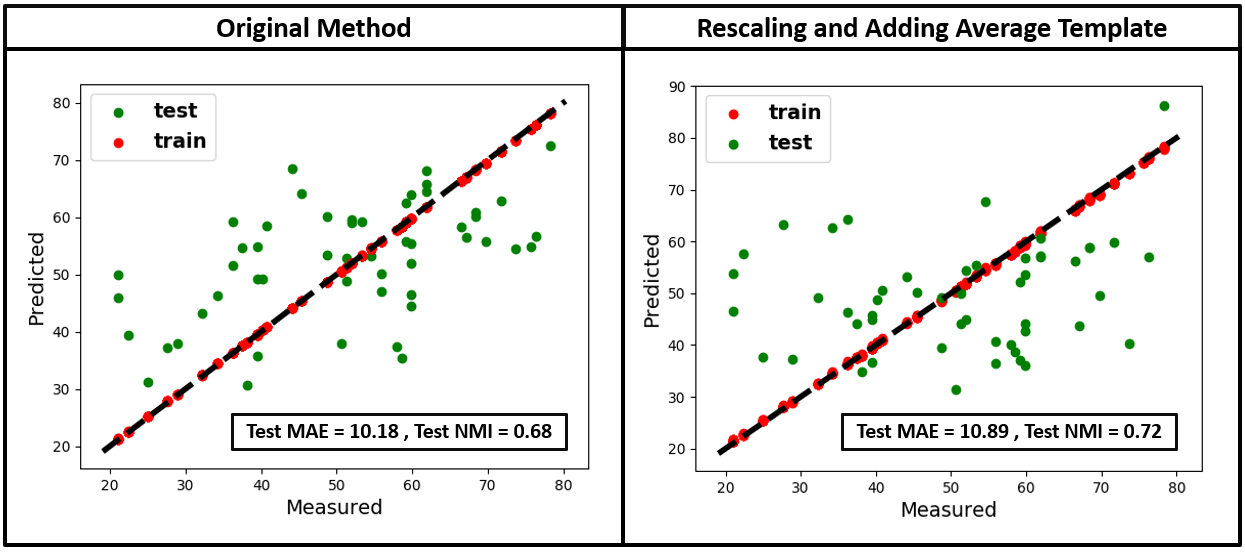}
   \caption{A performance comparison for Praxis prediction after modifying the objective according to Eq.~(\ref{eqn22:Eqn22}). \textbf{(L)} Original Method \textbf{(R)} After re-scaling and average template addition}\label{fig10:Fig10}
\end{figure}

 The green plots in Fig.~\ref{MitPar} illustrate the cross validated performance of the modified JNO framework from Eq.~(\ref{eqn23:Eqn23}). The operating point $\{\lambda_{1},\lambda_{2}\}$ for the modified framework is fairly consistent across the scores. Moreover, the green plots exhibit a larger stable range (highlighted in yellow) compared to the red plots. Accordingly, we identify the settings $\{\lambda_1 = 10$--$30$, $\lambda_2 = 0.08$--$0.6\}$ as the operating range for the modified JNO objective, which is roughly an order of magnitude larger than the original formulation and does not exhibit any kinks. Fig.~\ref{fig9:Fig9} and Fig.~\ref{fig10:Fig10} illustrate the best generalization performance for SRS and Praxis using the two algorithms.

\par Notice that the modified JNO has a slight tradeoff in regression performance at the expense of the gain in parameter stability. We highlight the importance of this exploration, as future applications of our work include applying our method to rs-fMRI and severity scores from a variety of neurological disorders. Our modified formulation provides additional flexibility in this sense, and extends the overall generalizability of our model.
\section{Discussion}
\par Our JNO model cleverly exploits the structure intrinsic to rs-fMRI correlation matrices through an outer product representation. The regression term further guides the basis decomposition to explain the group level and patient specific information. The compactness of our representation serves as a dimensionality reduction step that is related to the clinical score of interest, unlike the pipelined treatment commonly found in the literature. As seen from the results, our JNO framework outperforms a wide range of well established baselines from the machine learning and graph theoretic methods ubiquitous in fMRI analysis on two separate real world datasets.
\par We conjecture that the baseline techniques fail to extract representative patterns from the correlation data, and learn only the group level representation for the cohort. Consequently, they overfit the training set, despite sweeping the parameters across several orders of magnitude. Any patient level symptomatic and connectivity level differences are lost due to the restrictive pipelined procedure and the group level confounds. 
\par Our Joint Network Optimization Framework is agnostic to the choice of parcellation scheme. This was demonstrated by our additional experiments on the KKI dataset, where we chose the $246$ region Brainnetome parcellation to extract correlation matrices (Section~{\ref{MitiPara}}). We emphasize that our framework makes minimal assumptions on the data. Provided we have access to a valid behavioral and network similarity measure, this analysis can be easily adapted to other neurological disorders and even predictive network models outside the medical realm. This greatly broadens the scope of the method to numerous potential applications. 
\par Finally, notice that the training examples (red points) in Figs.~\ref{fig6:Fig6}$-$\ref{fig7:Fig7 ABIDE-SRS} follow the $\mathbf{x}=\mathbf{y}$ line nearly perfectly. Here, we explain this (potentially misleading) phenomenon in terms of the parametrizatization of our joint objective in Eq.~(\ref{joint_objective}).
\par
\begin{figure}[t!]    
    \centering
   \includegraphics[scale=0.36]{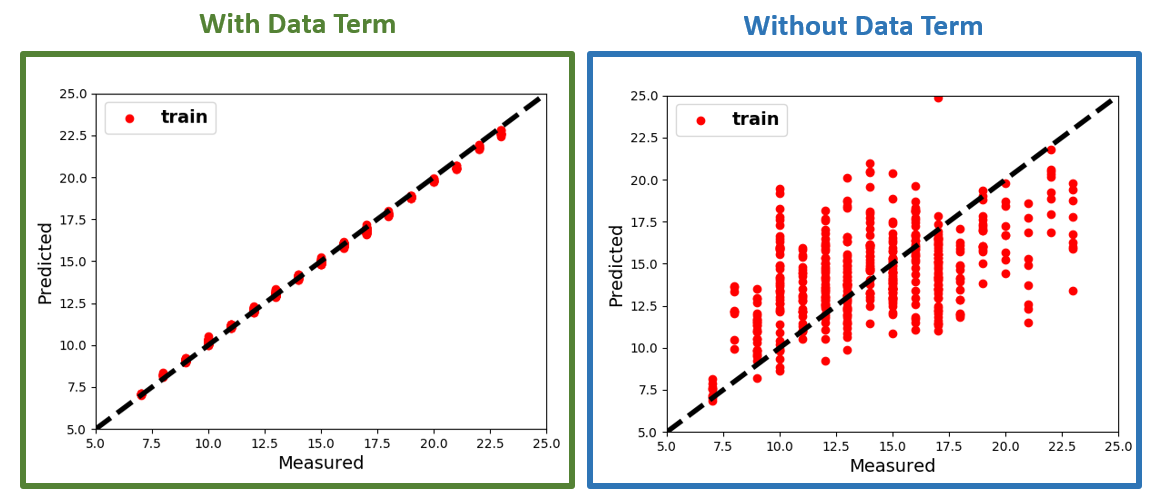}
   \caption{Prediction Performance of the JNO for ADOS on training data when \textbf{(L)} The data term is included in computing $\mathbf{c}_{n}$ \textbf{(R)} The data term is excluded from the computation of $\mathbf{c}_{n}$ }
   \label{OverP}
\end{figure}

 Recall that Section~\ref{pred_unseen} describes the procedure for calculating the coefficients for an unseen patient $\bar{\mathbf{c}}_{n}$ from the training solution set $\{{\mathbf{B}^{*}},{\mathbf{w}^{*}}\}$. Recall that we explicitly set the contribution from the data term in Eq.~(\ref{eqn5:Eqn5}) to $0$. Since the patient is not a part of the training set, the corresponding value of $\mathbf{\hat{y}}$ is unknown. In contrast, the training performance is computed based on the estimated coefficients $\mathbf{c}_{n}$, which have access to the severity scores. Here, we examine the effect of removing the severity information when calculating the coefficients for the training patients. In other words, we estimate the corresponding severity $\mathbf{y}$ excluding the ridge regression term. Accordingly, Fig.~\ref{OverP} highlights the differences in training fit with and without this term is not included in estimating $\mathbf{c}_{n}$. Notice that in the latter, the training accuracy has the same distribution as the testing points in Figs.~\ref{fig6:Fig6}$-$\ref{fig7:Fig7 ABIDE-SRS}. Taken together, we conclude that, the linear predictive term overparamterizes the search space of solutions for $\mathbf{c}_{n}$ to yield a near perfect fit. We use this observation to emphasize that the subnetworks and regression model learned by our JNO framework are capturing the underlying data distribution and not simply `overfitting' the training data.
\section{Conclusion}
We present an elegant matrix decomposition strategy to combine neuroimaging and behavioral data information. As opposed to generic prediction frameworks, our model directly captures key representative information from the correlation matrices. In contrast, conventional analysis methods dramatically fall short of unifying the two data viewpoints reliably enough to implicate predictive functional patterns in the brain. Our joint optimization framework robustly identifies brain networks characterizing ASD and provides a key link to quantifying and interpreting the clinical spectrum of manifestation of the disorder across patients. Moreover, our evaluation on two separate real world dataset supports the reproducibility of the framework.
\par We are working on a multi-score~\cite{d2019integrating} extension which can incorporate data from different behavioral domains. In the future, we will explore extensions of this model that learn a patient versus controls distinction in addition to predicting symptom severity. A potential extension of this model includes replacing the linear regression term in Eq.~(\ref{eqn5:Eqn5}) with its non-linear counterpart~\cite{d2019coupled}, thus providing us with the flexibility to model more complex decision functions which can better map the behavioral space.
\par Another avenue for refinement is to incorporate structural connectivity information in the form of anatomical priors~\cite{d2020deep}. Typically, structural modalities such as Diffusion Tensor Imaging (DTI) are used to define and track existing anatomical pathways in the brain. Incorporating this information into the network optimization model could be an important step towards unifying anatomical, functional and behavioural domains to better understand altered brain functioning in the context of neurological disorders such as Autism, ADHD, and Schizophrenia.
\par Our experiment on the KKI dataset using the Brainetome parcellation supports the observation that the method can be tuned with different connectivity and behavioral information, which is an added flexibility. Thus, it could be used to characterize the efficacy of behavioral therapies for neuropsychiatric disorders, as well as, for the development of patient-specific disease biomarkers. We believe that the all of benefits offered by our JNO framework could make it an important diagnostic tool for personalized medicine in the future. 
\paragraph{\textbf{Acknowledgements}}
This  work  was  supported  by the National Science Foundation CRCNS award 1822575 and CAREER award 1845430, the National  Institute  of Mental Health (R01 MH085328-09, R01 MH078160-07, K01 MH109766 and R01 MH106564), the National Institute of Neurological Disorders and Stroke (R01NS048527-08), and the Autism Speaks foundation.
\section*{Appendix A}
\label{Appd}
In this section, we derive the alternating minimization updates from Section \ref{Optim}:
\paragraph{\textbf{Optimizing $\mathbf{B}$ via Proximal Gradient Descent}} We first write out the optimization problem with respect to $\mathbf{B}$ when the estimates of $\{\mathbf{C},\mathbf{w}\}$ are held constant:
\begin{multline*}
\mathbf{B}^{k+1} = \argmin_{\mathbf{B}}{{\lambda_{1}{\vert\vert{\mathbf{B}}\vert\vert}_{1}}+{\sum_{n}}{\vert\vert{\mathbf{\Gamma}_{n}-\mathbf{D}_{n}\mathbf{B}^{T}}\vert\vert}_{F}^{2}}  \\
+ \sum_{n}{\Tr{\left[{\mathbf{\Lambda}_{n}^{T}({\mathbf{D}_{n}-\mathbf{B}\mathbf{diag}(\mathbf{c}_{n})})}\right]}}
\\ + \sum_{n}{{\frac{1}{2}}{\vert\vert{\mathbf{D}_{n}-\mathbf{B}\mathbf{diag}(\mathbf{c}_{n})}\vert\vert}_{F}^{2}} \ \ \ \ \ \ \ \ 
\\ \mathbf{B}^{k+1} = \argmin_{\mathbf{B}}{{\vert\vert{\mathbf{B}^{k}}\vert\vert}_{1}+ \frac{1}{\lambda_{1}}\mathcal{G}{(\mathbf{B}^{k})}} \ \ \ \  \ \ \ \ \ \ \ \ \ \ \ \ \ \ \ \ \ \ \ 
\end{multline*}
We see that the proximal gradient iteration is the solution to the following fixed point problem:
\begin{equation*}
\mathbf{0} \in {\frac{1}{\lambda_{1}}}\frac{\partial \mathcal{G}}{\partial \mathbf{B}} + \partial({\vert\vert{\mathbf{B}}\vert\vert}_{1})
\end{equation*}
The derivative of $\mathcal{G}$ with respect to $\mathbf{B}$, is computed as:
\begin{multline*}
\ \ \ \frac{\partial \mathcal{G}}{\partial \mathbf{B}} = \sum_{n}\left[{{2\left[{\left[\mathbf{B}\mathbf{D}_{n}^{T}-\mathbf{\Gamma}_{n}\right]\mathbf{D}_{n}}\right]-\mathbf{D}_{n}\textbf{diag}(\mathbf{c}_{n})}}\right] \\ +\sum_{n}{\left[{\mathbf{B}\textbf{diag}(\mathbf{c}_{n})}^{2}-\mathbf{\Lambda}_{n}{\textbf{diag}(\mathbf{c}_{n})}\right]}\ \ \ \ \ \ \ \ \ \ \ \ \
\end{multline*}
Given the fixed learning rate parameter $t$, the proximal update for $\mathbf{B}$ is easily computed as:
\begin{eqnarray*}
\ \ \ \ \ \ \ \ \ \ \mathbf{B}^{k+1} = \mathbf{sgn}(\mathbf{X})\circ(\mathbf{max}(\vert{\mathbf{X}}\vert-t,\mathbf{0})) \\  \mathbf{X} = \mathbf{B}^{k} - (t/\lambda_{1})\frac{\partial \mathcal{G}}{\partial \mathbf{B}} \ \ \ \ \ \ \ \ \ \ \ \ \ \ \ \ \
\end{eqnarray*}
This step first estimates a locally smooth quadratic model at each iterate $\mathbf{B}^{k}$ and applies a step of iterative shrinkage thresholding to the compute the local solution of $\mathbf{B}$. The resulting iterative algorithm is computationally efficient compared to the counterpart sub-gradient based descent methods and arrives at a good local solution for an appropriate choice of the learning rate.

\paragraph{\textbf{Optimizing $\mathbf{C}$ using Quadratic Programming}}
The objective is quadratic in $\mathbf{C}$ when $\mathbf{B}$, and $\mathbf{w}$ are held constant. Furthermore, the $\mathbf{diag}(\mathbf{c}_{n})$ term decouples the updates for $\mathbf{c}_{n}$ across patients. Each $\mathbf{c}_{n}$ is the solution to the a separate optimization problem of the following form:
\begin{multline*}
\mathbf{c}^{k+1}_{n} = \argmin_{\mathbf{c}_{n} \in \mathcal{R}^{K+}}\Tr{\left[\mathbf{\Lambda}^{T}_{n} ({\mathbf{D}_{n}-\mathbf{B}\mathbf{diag}(\mathbf{c}^{k}_{n})})\right]} +\lambda_{2}{{\vert\vert{\mathbf{c}^{k}_{n}}\vert\vert}^{2}_{2}} \\ + \frac{1}{2} {\vert\vert{\mathbf{D}_{n}-\mathbf{B}\mathbf{diag}(\mathbf{c}^{k}_{n})}\vert\vert}^{2}_{F}  + \gamma((\mathbf{c}_{n}^{k})^{T}\mathbf{w}-\mathbf{y}_{n})^2
\end{multline*}
Hence, we use $N$ quadratic programs (QP) of the form below to solve for the vectors $\{\mathbf{c}_{n}\}$ : 
\begin{equation*}
\frac{1}{2}{\mathbf{c}_{n}^{T}\mathbf{H}_{n}\mathbf{c}_{n}} + \mathbf{f}_{n}^{T}\mathbf{c}_{n} \ \ s.t. \ \ \mathbf{A}_{n}\mathbf{c}_{n} \leq \mathbf{b}_{n}
\end{equation*}
The QP parameters for our problem are given by:
\begin{eqnarray*}
\mathbf{H}_{n} = \mathcal{I}_{K} \circ (\mathbf{B}^{T}\mathbf{B}) + 2\gamma{\mathbf{w}\mathbf{w}^{T}}+ 2\lambda_{2}\mathcal{I}_{K} \ \ \ \ \     \\  \ \ \ \ \ \mathbf{f}_{n} = -2\left[\mathcal{I}_{K}\circ(\mathbf{D}_{n}^{T}+\mathbf{\Lambda}_{n}^{T})\mathbf{B}\right]\mathbf{1}  -2\gamma y_n\mathbf{w}; \ \  \\ \mathbf{A}_{n} = -\mathcal{I}_{K} \ \ \ \mathbf{b}_{n} = \mathbf{0}  \ \ \ \ \ \ \ \ \ \ \ \ \ \ 
\label{eqn11:Eqn11}
\end{eqnarray*}
The non-negativity constraint requires us to project the quadratic programming solution to the space of positive reals in $K$ dimensions for each $\mathbf{c}_{n}$ through $\mathbf{A}_{n}$ and $\mathbf{b}_{n}$. Since the Hessians $\{\mathbf{H}_{n}\}$ for our problem are positive definite, there exist polynomial time algorithms for solving the bound constrained QPs to the global optimum value. The decoupling of the $\{\mathbf{c}_{n}\}$ allows us to solve for each coefficient vector in parallel.
\paragraph{\textbf{Closed Form Update for $\mathbf{w}$}} The global minimizer of $\mathbf{w}$ is computed at the first order stationary point of the convex objective, which is:
\begin{eqnarray*}
\mathcal{J}(\mathbf{w}) = \lambda_{3}{\vert\vert{\mathbf{w}}\vert\vert}^{2}_{2} + \gamma{\vert\vert{\mathbf{C}^{T}\mathbf{w}-\mathbf{y}}\vert\vert}^{2}_{2} \ \ \\ \ \ \ \ \ \ \ \ \ \ \ \frac{\partial \mathcal{J}}{\partial \mathbf{w}} = 0 = 2 \lambda_{3} \mathbf{w} + 2 \gamma (\mathbf{C}\mathbf{C}^{T}\mathbf{w} - \mathbf{C}\mathbf{y})
\end{eqnarray*}
The closed form update can be expressed as:
\begin{equation*}
\mathbf{w} = (\mathbf{C}\mathbf{C}^{T}+ \frac{\lambda_{3}}{\gamma} \mathcal{I}_{K})^{-1}(\mathbf{C} \mathbf{y})
\end{equation*}
Thus, the ratio $\frac{\lambda_{3}}{\gamma}$ acts as a regularizer for the matrix inversion in our estimate, ensuring that the update for $\mathbf{w}$ is well defined at each iterate. 
\paragraph{\textbf{Optimizing the Constraint Variables $\mathbf{D}_{n}$ and $\mathbf{\Lambda}_{n}$}}
A closed form solution for the primal variables $\{\mathbf{D}_{n}\}$ can be obtained by setting their first derivatives to zero: 
\begin{multline*}
\frac{\partial {\mathcal{J}}}{\partial {\mathbf{D}_{n}}} = 0 = \mathbf{diag}(\mathbf{c}_{n})\mathbf{B}^{T} + 2\mathbf{\Gamma}_{n}\mathbf{B} \\ - \mathbf{\Lambda}_{n} - \mathbf{D}_{n}  -2 \mathbf{D}_{n}\mathbf{B}^{T}\mathbf{B} \ \ \ \ \ \  \ \ \ \ \ \ \ 
\end{multline*}
\begin{equation*}
\mathbf{D}_{n} = (\mathbf{diag}(\mathbf{c}_{n})\mathbf{B}^{T}+ 2\mathbf{\Gamma}_{n}\mathbf{B} - \mathbf{\Lambda}_{n})(\mathcal{I}_{K}+2\mathbf{B}^{T}\mathbf{B})^{-1}
\end{equation*}
The gradient ascent update on  $\{\mathbf{\Lambda}_{n}\}$ is as follows:  
\begin{eqnarray*}
\ \ \ \ \ \frac{\partial \mathcal{J}}{\partial \mathbf{\Lambda}_{n}} = \mathbf{D}_{n}-\mathbf{B}\mathbf{diag}(\mathbf{c}_{n})
\\ \mathbf{\Lambda}_{n}^{k+1} = \mathbf{\Lambda}_{n}^{k} + \eta_{k}\frac{\partial \mathcal{J}}{\partial \mathbf{\Lambda}_{n}} \ \ \ \ \
\end{eqnarray*}
Similar to the case of the coefficients $\mathbf{c}_{n}$, each of the $N$ pairs of updates $\{\mathbf{D}_{n},\mathbf{\Lambda}_{n}\}$ are decoupled from each other, and can be solved in parallel. Overall, the sets of $\mathbf{\Lambda}_{n}$ gradient ascent updates ensure that the respective set of constraints $\mathbf{D}_{n}= \mathbf{B}\textbf{diag}(\mathbf{c}_{n})$ is satisfied with increasing certainty at each iteration. The Augmented Lagrangian construct ${\vert\vert{\mathbf{D}_{n}-\mathbf{B}\textbf{diag}(\mathbf{c}_{n})}\vert\vert}^{2}_{F}$ prevents trivial Lagrangian $\mathbf{\Lambda}_{n}$ solutions. 

\bibliography{mybibfile.bib}

\end{document}